\documentstyle[preprint,aps,epsf]{revtex}
\tighten
\begin{document}

\draft

\title{
Three-Nucleon Force Effects in Nucleon Induced Deuteron Breakup: 
Predictions of Current Models (I)
}

\author{
J.~Kuro\'s-\.Zo{\l}nierczuk$^1$, H.~Wita{\l}a$^1$, 
J.~Golak$^{1,2}$, H.~Kamada$^3$, A.~Nogga$^4$, R.~Skibi\'nski$^1$, 
W.~Gl\"ockle$^2$,
}
\address{$^1$M. Smoluchowski Institute of Physics, Jagiellonian University,
 Reymonta 4,  30-059 Krak\'ow, Poland}
\address{$^2$Institut f\"ur Theoretische Physik II,
         Ruhr Universit\"at Bochum, D-44780 Bochum, Germany}
\address{$^3$Department of Physics, Faculty of Engineering, Kyushu Institute
of Technology \\ 1-1 Sensucho, Tobata, Kitakyushu 804-8550, Japan}
\address{$^4$Department of Physics, University of Arizona, Tucson, Arizona, 
85721, USA
}

\date{\today}
\maketitle

\begin{abstract}
An extensive study of three-nucleon force  
effects in the entire phase space of the nucleon-deuteron 
breakup process, for energies from above the 
deuteron breakup threshold up to 200 MeV, has been performed. 
3N Faddeev equations have been solved rigorously using the modern high 
precision nucleon-nucleon  potentials AV18, CD~Bonn, Nijm~I,~II and 
Nijm~93, and also adding 3N forces. We compare predictions 
for cross sections and various polarization observables when NN forces are 
used alone or when the $2\pi$-exchange Tucson-Melbourne  
3NF was combined with each of them. 
In addition AV18 was combined 
with the  Urbana IX 3NF and CD~Bonn with the TM' 3NF, which 
is a modified 
version of the TM 3NF, more consistent with chiral symmetry. 
Large but generally model dependent 3NF effects have been found in certain 
breakup configurations, especially at the higher energies, both for cross 
sections and spin observables. These results demonstrate the usefulness 
of the kinematically complete  
breakup reaction in testing  the proper structure of 3N forces.
\end{abstract}

\narrowtext

\section{Introduction}
\label{secIN}

The nucleon induced deuteron breakup reaction together with  elastic 
nucleon-deuteron (Nd) scattering have been considered for long 
as a valuable tool to test 
the three-nucleon (3N) Hamiltonian, in particular to shed light on 
the importance and 
structure  of a three nucleon force (3NF) in the potential energy 
of the 3N system~\cite{meyer,lect-note}. 
However, up to recently difficulties in solving 
rigorously 3N equations in the continuum for realistic forces 
prohibited clear statements in such a study. 
The rapid progress in supercomputer  
technology made it now possible to gain numerically exact solutions of 3N 
Faddeev equations for any nucleon-nucleon (NN) interaction, even 
including a 3NF~\cite{ref17}. 

Present day models of NN interactions, mostly phenomenological 
and/or based on a meson exchange picture, have achieved  a high degree of 
maturity. It was possible by adjusting their parameters to describe with 
high precision ($\chi^2 / data \approx 1$) NN data in the large energy 
range from threshold to about $350$~MeV~\cite{ref3,AV18,CDBONN,ref4}. 
These so called realistic potentials
 AV18~\cite{AV18}, CD~Bonn~\cite{CDBONN,ref4}, 
Nijm~I,~II, and Nijm~93~\cite{ref3}, 
are now extensively used in few-nucleon studies. Recent developments 
of chiral perturbation theory 
lead to NN potentials, where the one- and multi-pion exchanges are treated 
unambiguously according to chiral 
symmetry~\cite{ref5,ref6,ref7,ref8,ref9,Kaiser,ref10,ref11,Robi}. 
Therefore 
the theoretical uncertainties of meson exchange based models have been 
at least partially removed and this step puts the theory of nuclear forces 
on a more solid basis. First successful applications of these interactions 
in 3N and 4N systems have already demonstrated their power in interpreting 
and describing the data~\cite{ref12,Entem,Epel1,Epel2,Entem1}.   
In this paper, however, we shall only apply the above mentioned 
high-precision NN interactions. 

This present state of NN potential models, describing NN data perfectly, 
suggests now, when going to systems with more than two nucleons, 
to concentrate on the significance and properties of 
many-body force contributions to the nuclear Hamiltonian.  
 In case of three nucleons the first 
indication on the  
importance of 3NF contributions came from $^3$He and $^3$H bound state 
studies. All realistic NN potentials are unable to reproduce their 
experimental binding energies leading to a $^3$He and $^3$H underbinding of 
the order of $0.5 \div 1$~MeV~\cite{ref13}. This clear cut 
underbinding exists also 
for $^4$He where it amounts to $2 \div 4$~MeV~\cite{ref13,ref14,Nogga1}. 
Also for higher mass nuclei 
up to A=8, where stochastic techniques must be applied, realistic 
NN forces failed to provide the experimental binding 
energies~\cite{ref15,ref15a}. 

A natural step to explain this underbinding was to consider 3NFs
in the nuclear Hamiltonian. Presently, the most often used dynamical 
process is the $2\pi$-exchange between three nucleons. 
An important dynamical ingredient in that process is, as suggested  long time ago by the 
Fujita-Miyazawa 3NF~\cite{ref36}, an intermediate 
excited nucleon state $\Delta$. It was augmented 
later by further ingredients leading to the Urbana IX~\cite{ref19} and 
Tucson-Melbourne~(TM)~\cite{Coon} 
3NFs, which are mostly used in present day few-nucleon calculations. 
It was possible by properly adjusting the parameters of these 
3NFs to get essentially correct 3N and 4N binding energies. Also a fairly 
successful description of low energy bound state energies of up to 
A=8 nuclei resulted when adding 3NFs.
This was recently improved by adding further 3NFs
related to 
three-pion exchange with intermediate $\Delta$'s~\cite{newArgonne}. In the 
present exploratory investigation we shall not yet include this newest 
generation of 3NFs.

Though this first signal on 3NF effects, 
resulting from discrete states, is important
 an even  more sensitive and detailed investigation of 3NF properties can
be carried through  in scattering processes, where already a 
rich set of spin observables has been measured and further data are expected to 
come~\cite{Sekiguchi,HOmeyer,bodek,sagara}. 
From the theoretical side one can make exact
 predictions  for these observables  using
nuclear forces in all their complexities~\cite{ref17}. Experimentally 
one can access nowadays
 spin observables in elastic Nd scattering where in the
initial states the deuteron and/or  the nucleon is polarized and 
where in the final state  the polarizations of the outgoing particles can also 
be  measured~\cite{ref18,ref18a,ref18b,ref18c,ref18d,Ermisch,ref18e,ref18f,ref18g}. 
This together with cross sections leads to a very rich 
spectrum of observables in Nd elastic scattering
and the Nd breakup process. Such a set of 
observables will be a solid basis to test the 3N Hamiltonian. Using 
available model
 Hamiltonians one can provide guidance to select specific
observables and energies which are most appropriate to see 3NF properties. 
 In~\cite{ref21a} such a study has been performed for elastic Nd scattering. 
It was shown that the current 3NF models exhibit a lot of effects and more 
data are 
needed to provide constraints on the theoretical models of 3NF´s. 
First data sets~\cite{Sekiguchi,ref18,ref18c,ref18d,Ermisch} 
in elastic Nd scattering showed both, successes 
and failures of present  day 3NF models combined with the realistic NN 
forces. It is the aim of this paper to probe the cross section and 
several spin observables of the kinematically complete Nd breakup process 
against sensitivity to 3NFs combined with the high-precision NN 
interactions. 
In a following paper (II) already existing data will be compared to 
those model 
predictions.
 
For the convenience of the reader we  briefly review in Section~\ref{secII} 
our theoretical formalism for the 3N continuum and display  
the 3NF models 
 which are presently in vogue and which we use.
Our predictions for quite a few breakup observables based on various 
nuclear force combinations are presented in Section~\ref{secIII}. 
While in this 
Section we investigate the full 3N breakup phase space we concentrate on 
a subset of breakup configurations in Section~\ref{secIV} according to 
the experimental restrictions imposed in an ongoing experiment at KVI~\cite{bodek,kistryn:01}. 
We conclude in Section~\ref{secV}. 

\section{Theoretical Formalism and 3NF Models}
\label{secII}

We refer to~\cite{ref17} for a general overview on 3N scattering and 
specifically for  
our way to formulate
it. The inclusion of 3NF´s follows \cite{ref21}. 
This is a direct
generalization of what is being used for the 3N bound state~\cite{ref22}. 
We define an amplitude $T$ via our
central Faddeev-like equation:
\begin{equation}
\label{eqT}
T \ = \ t \, P \, \phi \ 
+ \ ( 1 + t G_0 ) \, V_4^{(1)} \, ( 1 + P ) \, \phi \
+ \ t \, P \, G_0 \, T \
+ \ ( 1 + t G_0 ) \, V_4^{(1)} \, ( 1 + P ) \, G_0 \, T
\end{equation}
The initial channel state $\phi$  is composed of a deuteron and a
momentum eigenstate of the projectile nucleon. 
The two-body  $t$-operator is denoted by $t$, 
the free 3N propagator by $G_0$ and $P$ is 
the sum of a cyclical and an anticyclical 
permutation of three particles. 
The 3N force $V_4$ can always be decomposed into 
a sum of three parts:
\begin{equation}
\label{eqV4}
V_4 = V_4^{(1)} + V_4^{(2)} + V_4^{(3)} ,
\end{equation}
where $V_4^{(i)}$ is symmetrical under the exchange of the nucleons $j k$ 
with $j \ne  i \ne k$.
As seen in Eq.~(\ref{eqT}) only 
one of the three parts occurs explicitly, the others via the permutations
contained in $P$. 

The physical breakup amplitude is given via
\begin{equation}
U_0 = ( 1+ P) T .
\label{eqU0}
\end{equation}

The Faddeev-like integral equation~(\ref{eqT}) has the nice property 
that the terms resulting by iteration and inserted into Eq.~(\ref{eqU0}) 
yield  
the multiple scattering series, 
which gives  transparent insight into the reaction
mechanism. 

The definition of the various spin observables can be found 
in~\cite{ref17,ref24,Ohlsen2}. 
We shall encounter nucleon and deuteron vector analysing powers $ A_y$   
and $A_y^{(d)}$,  where in the initial state 
either the nucleon  or the deuteron  is polarized. 
Further, the deuteron can
be tensor polarized in the initial state 
leading to the three tensor analysing powers $T_{2k}$
$(k = 0,1,2)$ or the corresponding cartesian analyzing powers $A_{ij}$ 
$(i, j=x, y, z)$. 

We concentrate in this study on the kinematically complete 
3N breakup. The final 
3N state of this  reaction is specified by nine 
momentum components. In such an 
experiment the energies and directions (polar and 
azimuthal angles $\theta_i$ and $\phi_i$) of two outgoing nucleons 
$(i=1,2)$ are measured. This together with restrictions imposed by energy 
and momentum conservation lead to a correlation of the 
measured energies 
$E_1$ and $E_2$, causing them to lie on a kinematical curve (S-curve) 
in the $E_1 - E_2$ plane. 
 All observables will be  presented 
as a function of the arc length S related to a point along this kinematical curve. 
It is a matter of convention to choose the location of S$=0$ 
on the S-curve and we use the one defined in Ref.~\cite{ref17}.

As NN forces we use the five 
 realistic NN interactions mentioned in the introduction. They are combined 
with 3NF models. The $2\pi$-exchange TM model~\cite{Coon}
has been around for quite some time and  
 is based on a low momentum expansion of the $\pi-N$ off- (the mass-) 
shell scattering amplitude.
It has the following form:
\[
V_4^{(1)} = { 1 \over {( 2 \pi )^6} } \, 
            { g^2_{\pi N N} \over { 4 m_N^2} } \,
          {  { {\vec \sigma}_2 \cdot {\vec Q} } \over {\vec{Q}^{\, 2} + m_\pi^2} } \,
          {  { {\vec \sigma}_3 \cdot {\vec {Q '}} } \over {\vec{Q '}^{\, 2} + m_\pi^2} } \,
         H \left( \vec{Q}^{\, 2} \right) \, H \left( \vec{Q '}^{\, 2}  \right) \times
\]
\begin{equation}
\times \left\{ \vec{\tau}_2 \cdot \vec{\tau}_3 \left(  a \,   + \,  b \vec{Q} \cdot \vec{Q '} \,
                                                + \, c ( \vec{Q}^{\, 2} +  \vec{Q '}^{\, 2} ) \right)
               \ + \ d ~i \vec{\tau}_3 \times \vec{\tau}_2 \cdot  \vec{\tau}_1 \,
                  \vec{\sigma}_1 \cdot {\vec Q} \times \vec{Q '}   \right\} .
\label{eqTM3NF}
\end{equation}

The elements of the underlying Feynman 
diagram are obvious: the two-pion propagators depending on
the pion momenta $\vec{Q}$  and $\vec{Q '}$,  
the strong form factors combined into the two $H$-functions 
 and most importantly the parameterization
of the $\pi N $ amplitude inside the curly bracket which is combined 
with the isospins  $\vec{\tau}_2$  and $\vec{\tau}_3$ of
the two accompanying nucleons. 
The $H$-functions are parameterized as: 
\begin{equation}
 H \left( \vec{Q}^{\, 2} \right) \, = \,
\left(  { {\Lambda^2 - m_\pi^2} \over {\Lambda^2 + \vec{Q}^{\, 2}} }  \right)^2
\label{eqH}
\end{equation}
In what we denote by the TM 3NF we use the original parameters
$a=1.13/m_{\pi}$, $b=-2.58/m_{\pi}^3$, $c=1.0/m_{\pi}^3$, 
$d=-0.753/m_{\pi}^3$ where  $m_{\pi}=130.6$ MeV~\cite{ref25}.
The still very demanding computer resources prevented us to use the 
recently updated values\cite{Coon1} which, however, differ only slightly  from the one we use and 
 we do not expect significant changes. The $b$- and $d$-terms are mostly determined  by an  
  intermediate  $\Delta$ in a static 
approximation. The $a$- and $c$-terms are related to $S$-wave $\pi$N scattering. 
The cut-off parameter $\Lambda$ is used 
to adjust the $^3$H  binding 
energy separately for different NN forces~\cite{ref13}.
 In units of the pion mass $m_{\pi}$ we find the $\Lambda$ to be: 
 4.856, 5.215, 
5.120, 5.072, and 5.212 when the TM 3NF is combined with 
CD~Bonn, AV18, Nijm~I,~II, and Nijm~93, respectively.  

Of course in a meson exchange picture 
additional processes should be added containing 
different mesons  like $\pi - \rho$,
$\rho - \rho$, etc.; different intermediate 
excited states might also play a role~\cite{ref16}. To some extent
3NF models with respect to those extensions have already been developed 
and applied~\cite{ref27,ref28,ref29,ref30,Wita}. 
Our adjustment of $\Lambda$ is a very rough manner to take 
other processes into account. 
Further studies 
incorporating these additional 3NFs 
should be performed both for elastic Nd 
scattering and the breakup process. 

The parameterization of the TM 3NF has been criticized,  
since it violates chiral
symmetry~\cite{ref31,ref32}. A form 
consistent with chiral symmetry (though still incomplete  
to that order in the appropriate power counting) is 
obtained by keeping the long range part of the $c$-term leading to a changed parameter
 $a'\equiv a - 2m_{\pi}^2 = -0.87/m_{\pi}$~\cite{ref31,ref32} 
and dropping the short range part. This essentially means a change of 
sign for $a$. 
 This form will be called  TM' later on. The corresponding $\Lambda$-value, 
when TM' is used with the CD~Bonn potential, is $\Lambda = 4.593$.

The two-meson exchange 3NF has been also studied 
by Robilotta {\em et al.}~\cite{ref33}
leading to the Brazilean
3NF. It is similar to the one of TM and also the results gained 
for 3N observables~\cite{ref34}
are similar to the ones for the TM 3NF. In this paper
we do not take that force into account.
Instead we included the Urbana IX 3NF~\cite{ref19}, which 
is intensively used in the Urbana-Argonne
collaboration. 
 That force is based on the old Fujita-Miyazawa ansatz~\cite{ref36} 
of an intermediate 
$\Delta$ occurring in the two-pion exchange and
is augmented by a spin and isospin 
independent short range piece. It has the form
\[
V_4^{(1)} = A_{2 \pi} \, \left[ \{ X_{12} ,  X_{13} \} \,
               \{ \vec{\tau}_1 \cdot 
\vec{\tau}_2 , \vec{\tau}_1 \cdot \vec{\tau}_3 \} +\,
  \, \frac14 \, [ X_{12} ,  X_{13} ] \, 
    [ \vec{\tau}_1 \cdot \vec{\tau}_2 , 
\vec{\tau}_1 \cdot \vec{\tau}_3 ] \, \right] + \ 
\]
\begin{equation}
+ \ U_0 \, T_\pi^2 (r_{12}) \, T_\pi^2 (r_{13}) , 
\label{eqURBANA3NF}
\end{equation}
 where $\{,\}$ and $[,]$ are anticommutator and commutator brackets and $ X_{ij}$ is defined as: 
\begin{equation}
 X_{i j} = Y_\pi (r_{i j}) \,
\vec{\sigma}_i \cdot \vec{\sigma}_j \ + \ T_\pi (r_{i j}) \, S_{i j} 
\label{eqXIJ}
\end{equation}
Here $\sigma_i$ is a nucleon spin operator, 
$S_{ij}$ is the  standard tensor force,
the $T(r_{ij})$ and $Y(r_{ij})$ are  functions of the  distance $ r_{ij}$ 
between nucleons $i$ and $j$, that are  
associated with the Yukawa and  tensor part of 
the one-pion-exchange interaction
\begin{equation}
Y_\pi (r) \, = \, 
{ e^{- m_\pi r} \over {  m_\pi r } } \, \left( 1 - e^{- c r^2} \right) ,
\label{eqYPI}
\end{equation}
\begin{equation}
T_\pi (r) \, = \, \left[ 1 + { 3 \over 
{m_\pi r} } + { 3 \over {(m_\pi r)^2} } \right] \,
{ e^{- m_\pi r} \over {  m_\pi r } } \, \left( 1 - e^{- c r^2} \right)^2 ,
\label{eqTPI}
\end{equation}
quenched via a short-range cut-off function 
$(1-e^{-c r^2})$ with $c$=2.1~fm$^{-2}$~\cite{carlson:83}. 
The parameters  $A_{2\pi}$ and  $U_{0}$ were chosen to reproduce 
the experimental $^3$H binding energy 
and to  obtain 
the empirical equilibrium density of nuclear matter when 
this 3NF model is used with the AV18 potential. 
Their values for model IX are 
$A_{2\pi}=-0.0293$~MeV and $U_{0}=0.0048$~MeV~\cite{ref19}. 
Since we work in momentum space using a partial wave expansion, the form 
given in Eq.~(\ref{eqURBANA3NF}) has to be
rewritten.  This has been done in~\cite{ref21a}. 

Since there is no apparent consistency of the mostly phenomenological 
realistic NN forces and the 3NF models, we test various combinations thereof.
In all cases, however, we require that the particular force 
combination  should reproduce 
the experimental triton binding energy. Some of the 3N observables 
scale with the triton binding energy~\cite{wit99}. The adjustment 
to the triton binding energy has the advantage that our investigation 
is not misled by these scaling effects.  

With respect to the intermediate $\Delta$ one should say that 
very likely the static approximation
is not justified and the $\Delta$ should be allowed to propagate 
like  the nucleon. This has been
pursued intensively, for instance, by the Hannover group~\cite{ref38} 
and their recent work has 
been also devoted to the 3N continuum~\cite{ref39}. 

In view of all that, it is quite 
clear that our present study is not at all 
complete but can at least provide some insight into 
 effects specific 3NF models  generate for breakup 
observables.

\section{Predictions of 3NF Effects}
\label{secIII}

Since we would like to cover a wide range of incoming nucleon 
energies, ranging  from just above the Nd breakup threshold up to 200 MeV, it 
is necessary to take a sufficient number of partial waves into account to 
guarantee converged solutions of the Faddeev equations. In all presented 
calculations we went up to the two-nucleon subsystem total angular 
momentum $j_{max}=5$. This corresponds to a maximal number of 142 
 partial wave states (channels) in the 3N system. We checked that the 
convergence has been achieved by looking at the results obtained for 
$j_{max}=6$, which increases the number of channels to 194. This convergence 
check refers to a calculation without a 3NF.  The inclusion of 3NFs 
has been carried through for all total angular momenta of the 3N system up to 
$J=13/2$. These high angular momenta are required at the higher energies 
$\ge 100$~MeV. The longer ranged 2N interactions require states up to 
$J=25/2$ at the higher energies in order to get converged results.

A phenomenological criterion for 3NF effects is that they lie outside of 
the spread of the five realistic NN
force predictions only. This spread will be indicated 
by a band (called the  ``2N'' band). 
Unfortunately
we cannot include the $pp$ Coulomb force effects. 
 At the higher energies, however, 
those effects should
be small. Also in case of AV18 we do not take the various 
electromagnetic corrections into
account, which leads among possibly other effects to a slightly 
wrong deuteron binding energy ($E_b = 2.242$ instead of $2.225$~MeV). 
This shifted $E_b$-value, however, has only a small 
effect on our results, which is mostly 
of kinematical origin. Also the NN-phase shifts, obtained without 
those additional terms, differ only slightly from the standard ones. 
The kinematical effect also leads to a slight shift in the position of 
the S-curve. 

To investigate 3NF effects we
combine 
the TM 3NF with the five NN forces forming a second band (called the ``2N+TM'' band). 
In all cases the cut-off value 
$\Lambda$ in Eq.~(\ref{eqH}) has been adjusted separately for
each NN force to the $^3$H binding energy~\cite{ref26} 
as given in the previous Section. Since that interplay 
is a purely phenomenological step, the outcome is theoretically 
not under control 
and we combine all the results into a second band.
In addition, we want to compare the TM 3NF and the
modified  TM', which is more consistent with chiral symmetry. 
 We combine TM' with CD~Bonn and display the results  as a dashed curve. 
 Finally we compare
the AV18 with the  Urbana IX 3NF what is presented as a solid line.
 There are clearly 
more combinations possible but we felt this to be sufficient 
to get an orientation 
on the magnitudes of the effects.

\begin{figure}[h!] 
\leftline{\mbox{\epsfysize=11.0cm \epsffile{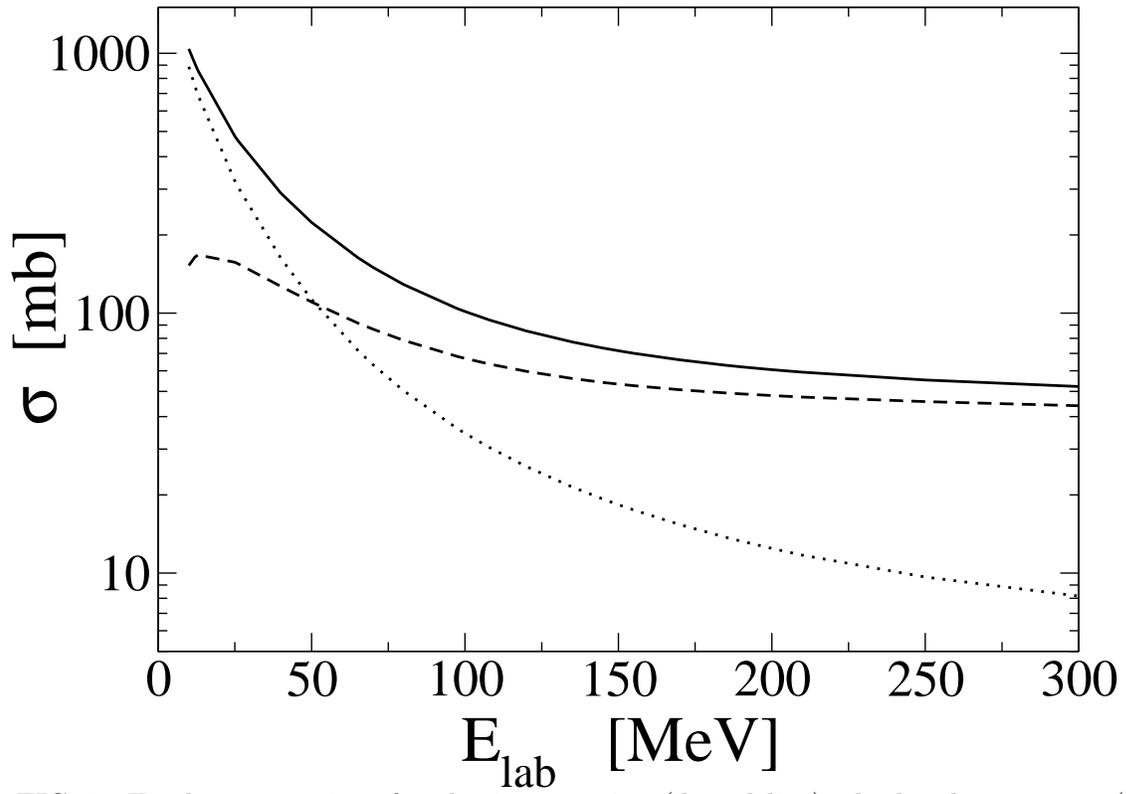}}} 
\caption[]{Total cross sections  for
elastic scattering  (dotted line), the breakup  process  (dashed line)
and their sum (solid line) as predicted by the CD Bonn potential.}
\label{fig:totelbr}
\end{figure}

Before starting to investigate  exclusive breakup configurations with large 
3NF effects, it is interesting to look at the elastic 
scattering and breakup contributions to the total cross section for 
neutron-deuteron ($nd$) scattering. 
 Fig.~\ref{fig:totelbr} shows the total cross 
section $\sigma_{tot}$, predicted 
by the CD~Bonn potential,  
as a function of the incoming neutron laboratory energy up to 300~MeV together 
with the contributions coming from elastic scattering and the 
breakup processes. 
It is seen that the total elastic cross section 
decreases rapidly with increasing energy of the incoming neutron.
At about 50~MeV the breakup contributes already $\sim$50\% to 
the total cross section. 
At higher energies it dominates the total cross section and  at about 200~MeV 
the elastic scattering contributes 
only $\sim$20\% to the total  cross section.  
This picture does not depend on the choice of the  NN potential used: 
the theoretical results for the $nd$ total elastic and breakup cross sections 
are stable under the exchange of the  realistic 
NN potentials. The differences are 
of the order of 1\%.

\begin{figure}[htbp]
\leftline{\mbox{\epsfysize=11.0cm \epsffile{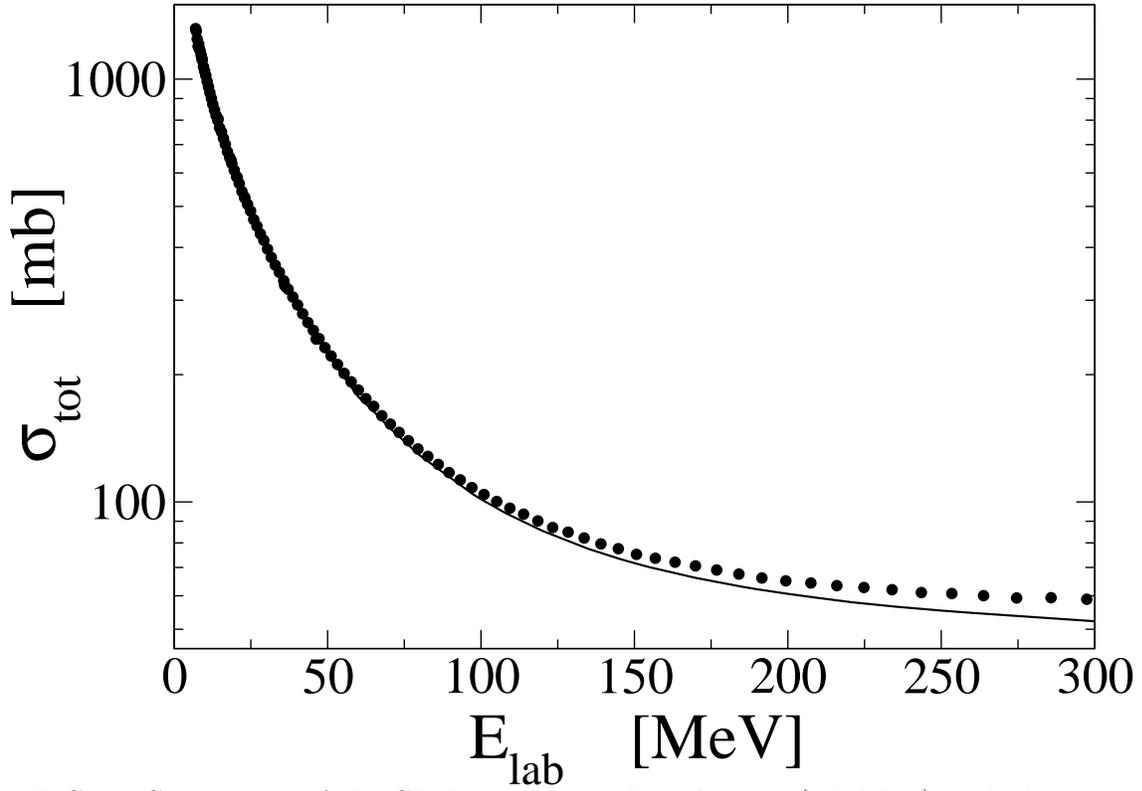}}}
\caption[]{Comparison of the CD Bonn potential 
predictions (solid line) with the experimental data (full dots)~\cite{Abfalte} for the 
total $nd$ cross sections $\sigma_{tot}$.}
\label{fig:allpot-totcross}
\end{figure}

\begin{figure}[htbp]
\leftline{\mbox{\epsfysize=11.0cm \epsffile{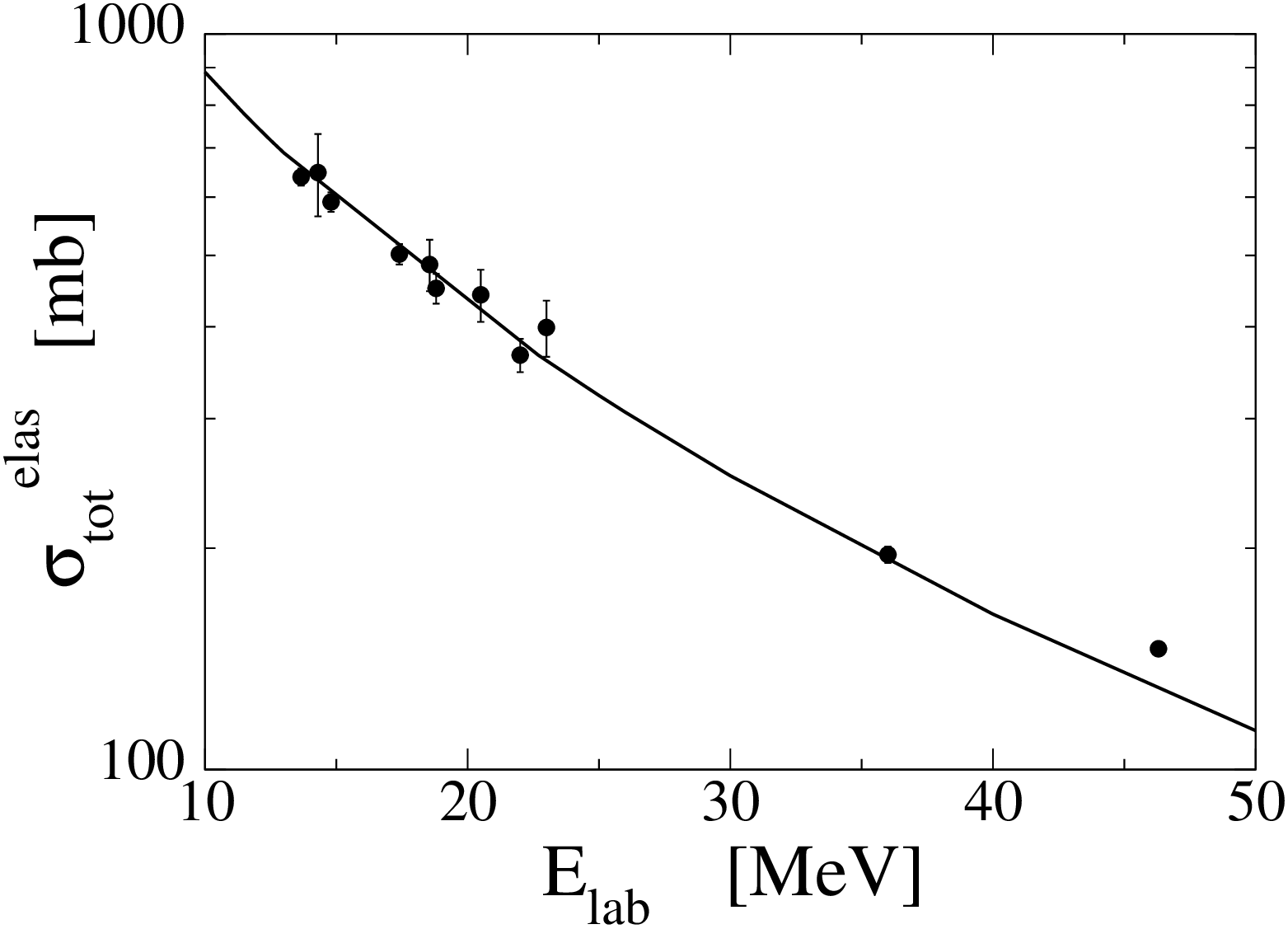}}}
\caption[]{Comparison of the CD Bonn potential 
predictions (solid line) with the experimental data (full dots)~\cite{seag72,paul75,holm69,catr61} 
for  the total elastic cross section $\sigma_{tot}^{elas}$.}
\label{fig:allpot-totcross-elas}
\end{figure}

\begin{figure}[htbp]
\leftline{\mbox{\epsfysize=11.0cm \epsffile{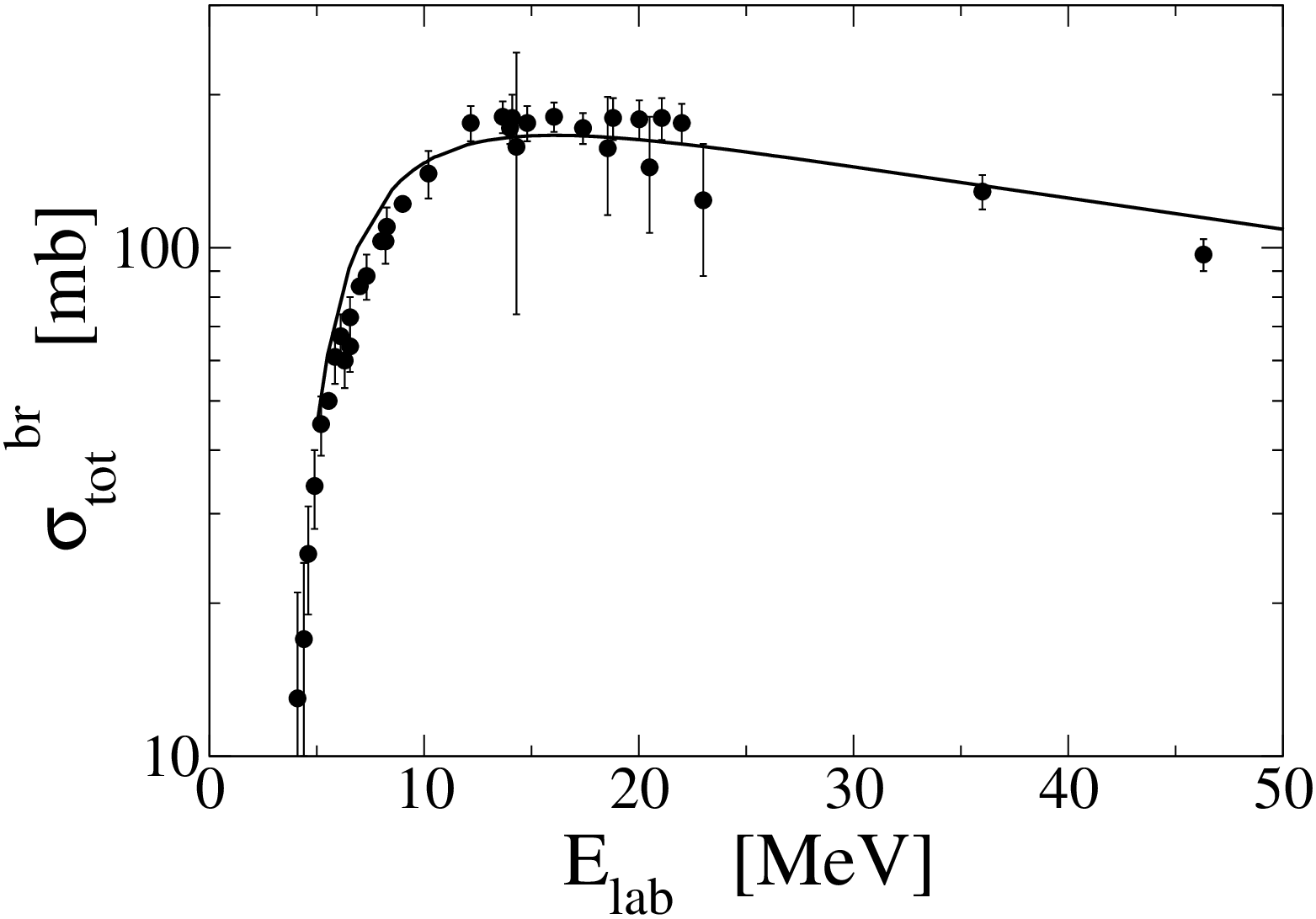}}} 
\caption[]{Comparison of the CD Bonn potential 
predictions (solid line) with the experimental data (full dots)~\cite{seag72,paul75,holm69,catr61} for the total breakup cross section $\sigma_{tot}^{br}$.}
\label{fig:allpot-totcross-br}
\end{figure}

In Figs.~\ref{fig:allpot-totcross}-\ref{fig:allpot-totcross-br} 
we compare the  theoretical predictions
with the experimental data.   
There is a good agreement between the ``pure'' NN  theory  and
 the data up to $\sim$50~MeV for all three total cross sections. 
At higher energies the data exist only for $\sigma_{tot}$ and 
the theory  begins to underestimate the data (about 2\%  at around 100~MeV
and about 13\% at $\sim$300~MeV)~\cite{Abfalte,Wita1}.

\begin{figure}[htbp]
\leftline{\mbox{\epsfysize=11.0cm \epsffile{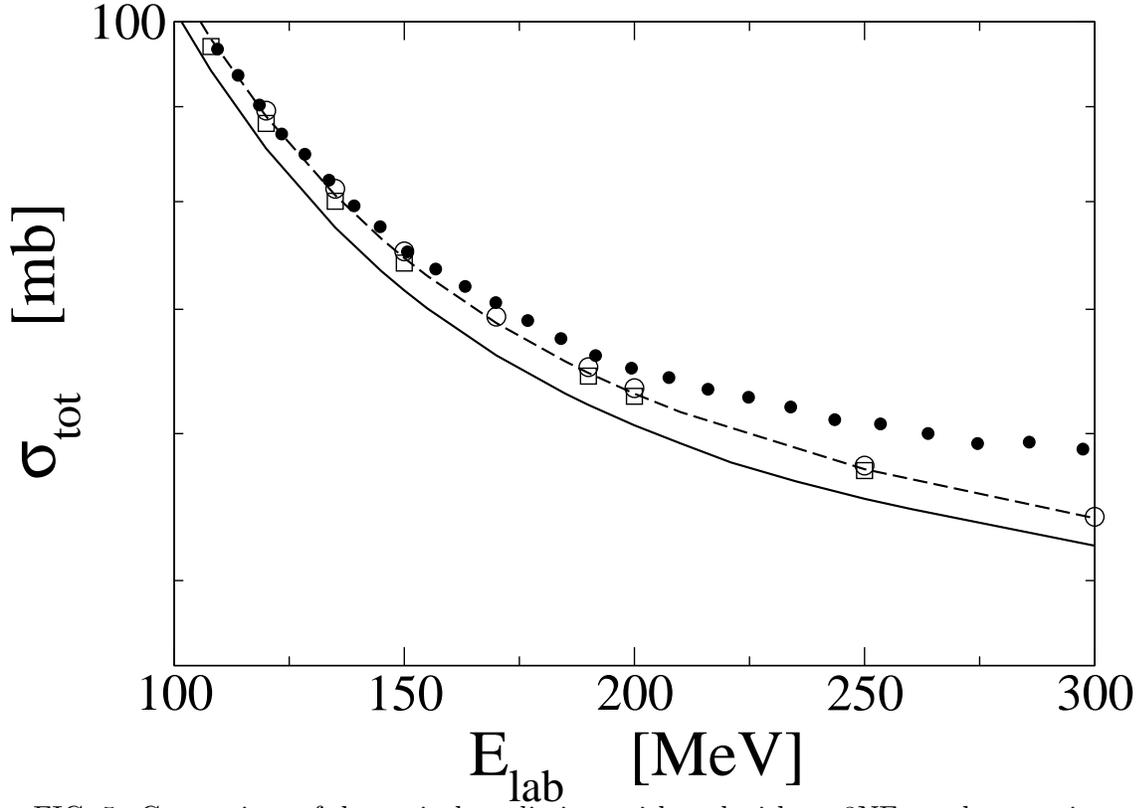}}}
\caption[]{Comparison of  theoretical predictions  with and without  
3NFs to the experimental data (full dots)~\cite{Abfalte} for the $nd$ total  cross section. 
The solid and dashed lines are the CD Bonn and CD Bonn+TM 3NF predictions, 
respectively. The open squares are the results for AV18+Urbana~IX
and the circles for CD Bonn+TM'.}
\label{fig:totcross-3nf}
\end{figure}

In order to see the influence of 3NFs, we calculated  the 
total cross section for $nd$ scattering combining some of the NN potentials
and 3NF models as described above.
We see, adding a 3NF increases the value of the different 
total cross sections and shifts 
the theoretical predictions 
closer to the data (it is clearly visible for $E_{lab} \gtrsim 100$~MeV in
Fig.~\ref{fig:totcross-3nf}). The predictions obtained with  
different 3NF models are
 practically  the same.

\begin{figure}[htbp]
\leftline{\mbox{\epsfysize=19.0cm \epsffile{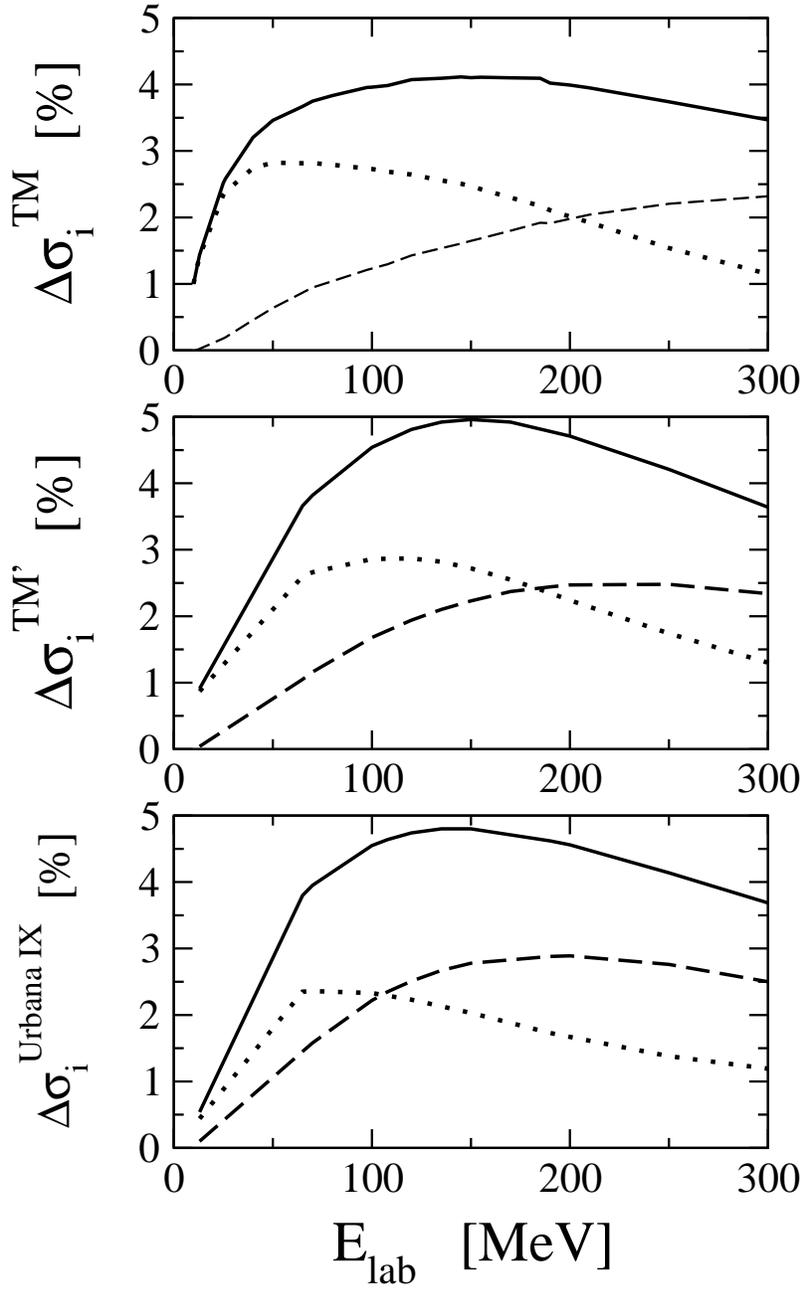}}}
\caption[]{3NF effects in the $nd$ total cross section (solid line),
the total elastic scattering cross section (dotted line), and the 
total breakup cross section (dashed line), expressed as a percentage 
change (see Eq.~\ref{crostot-delta}) with respect to the total cross section as 
obtained with the NN 
interactions alone: CD Bonn+TM (top), 
CD Bonn+TM' (middle), and AV18+Urbana~IX (bottom).}
\label{fig:diff-totcross}
\end{figure}

In order to quantify the effect of a 3NF on the total 
cross sections we
calculated its relative contributions  in the corresponding processes as:
\begin{equation}
\Delta \sigma^{3NF}_{i} =
\left |\frac{\sigma_{i}^{(2N)}-\sigma_{i}^{(2N+TM)}}
{\sigma_{tot}^{(2N)}}\right | \times 100\%
\label{crostot-delta}
\end{equation}  
where $\sigma_i \equiv \sigma_{tot}, \sigma_{tot}^{elas}, \sigma_{tot}^{br}$.
As  can be seen from Fig.~\ref{fig:diff-totcross},  the 3NFs
 enhance the $nd$ total  cross section only
by about up to 5\%. This enhancement is practically  
independent from  the particular 3NF used  
and for energies above 100~MeV changes only slightly with energy. 
Due to the drastically diminishing contribution of the 
total elastic cross section 
most of the 3NF effects at higher energies must occur 
in the breakup process. Since the total 
breakup cross section is dominated by the quasi free scattering, 
for which  3NF effects are negligible~\cite{ref17,phdkuros}, it follows, 
that at higher energies 
  3NF effects must be found
in other breakup configurations, possibly just in some specific geometries. 
It is one of the aims of this paper to investigate this conclusion in detail.

As can be seen from Fig.~\ref{fig:totcross-3nf}, addition of  a 3NF 
only partially removes the discrepancy 
between data and theory. 
At higher energies one has to expect that relativistic effects might appear.
A generally accepted framework for carrying out relativistic 3N  
calculations does not yet exist. An estimation of these effects 
due to  proper relativistic treatment of the kinematics  
while  keeping the (non-relativistic) scattering amplitudes unchanged 
was done  in~\cite{Wita1}.  
It was found, that such 
relativistic effects  enhance the $nd$ total  cross section 
by about 7\% at 250~MeV 
while being much smaller at lower energies. 
These corrections bring the theoretical 
predictions for the $nd$ total  cross section much closer to the 
data.

In the following  we restrict  the study of  3NF effects 
to the  
 $nd$ breakup reaction. Contrary  to elastic scattering, 
where the description of the two-body   final state requires 
only one variable (e.g. the c.m.  angle $\theta$), the 3N final 
state of the breakup process is  kinematically much more complex and 
requires five parameters to be described completely. 
The possibility to choose very many specific kinematical geometries and energy 
distributions for the three outgoing nucleons allows to probe 
the 3N Hamiltonian in greater detail. 

Our goal is to scan the full breakup 
phase space for the configurations which  exhibit 
large  3NF effects in some observables. 
In order to cover the broad 
range of incoming energies, we took, as representatives, the 
following laboratory energies: 13, 65, 135, and 200 MeV. 

As a first exploratory study, we calculated 
the five-fold differential 
cross section $\sigma \equiv \frac{d^5\sigma}{dS d\Omega_1 d\Omega_2}$, 
and the analyzing powers  
$A_{y}$, $A_{x}$, $A_{z}$, $A_{y}^{(d)}$, $A_{yy}$, $A_{xz}$, $A_{xx}$
using the  CD~Bonn NN potential with and without  the TM 3NF.
The observables were  tabulated  as a function of $E_{1}, E_{2}, 
\theta_{1}, \theta_{2}$ and $\phi_{12}\equiv \phi_1 - \phi_2$, varying  
$\theta_{i}$ in steps of $5^\circ$,   $\phi_{12}$ 
in steps of $10^\circ$ and taking a step along the S-curve, 
$\Delta~S(E_1,E_2)=1$~MeV, over the full  phase space. 

In order to check if 
the chosen grid  over the full phase space was sufficiently dense 
we calculated:
\begin{equation}
\sigma_{tot}^{scan} \equiv \sum_{i=1}^N 
\left (
\int d\Omega_1
d\Omega_2 
dS
\frac{d^5\sigma}{d\Omega_1 d\Omega_2 dS}\right )_i
\label{eq:scan}
\end{equation}
where $N$ is the number of the calculated configurations.
In Tab.~\ref{tab:br-scan} the values of $\sigma_{tot}^{scan}$ are  compared 
with  the total breakup cross section  $\sigma_{tot}^{br}$.
The difference between 
$\sigma_{tot}^{br}$ and $\sigma_{tot}^{scan}$ is smaller than  5\%,
 what gives a strong support that our grid  is sufficiently dense.

In order to discover  interesting cases  
we  looked for configurations with the largest differences 
between  observables predicted  with the 
CD~Bonn alone  and  when  the  TM 3NF has been included. 
The influence of the TM  3NF 
can be quantitatively defined by the relative 
difference  $\delta O $ of the observable 
$O=\sigma$, $A_{y}$, $A_{x}$, $A_{z}$, 
$A_{y}^{(d)}$, $A_{yy}$, $A_{xz}$, $A_{xx}$
at a given phase space point 
$(E_{1}, E_{2},\theta _{1}, \theta _{2},\phi_{12})$ 
when the TM 3NF is switched on 
\begin{equation}
\delta O (E_{1}, E_{2},\theta _{1}, \theta _{2},\phi_{12}) 
\equiv 
\left|
\frac{O^{(2N)}-O^{(2N+TM)}}{O^{(2N)}}\right| \times 100\%
\label{eqn:deltao}
\end{equation} 
Here ${O}^{(2N)}$ and ${O}^{(2N+TM)}$ are the   
predictions for the  particular observable $O$ 
   using the CD~Bonn potential alone and including in 
addition  the TM 3NF. 
In the case of analyzing powers, 
due to experimental limitations 
only kinematical points  have been considered for which 
$|A_{i}^{(2N)}|\ge 0.05$. 

In order to get  insight into the limitations imposed by the experimental 
conditions  in detecting interesting cases we performed the 
search in three steps. 
First, for each observable $O$ at each energy we searched over the 
entire grid of the phase space
for those points where $\delta O$ achieves its maximum value $\Delta O$: 
\[ 
        \Delta O \equiv \max [\delta O(E_{1}, E_{2},
\theta _{1}, \theta _{2},\phi_{12})] 
\]
The resulting $\Delta O$'s are presented 
in the rows labeled {\em all} in Tab.~\ref{tab:max4scan}.
Second, bearing in mind that we look for  effects 
that should be  experimentally accessible, we searched for the maximum 
of $\delta O$ 
only at phase space points for which the  cross section 
$\sigma^{(2N)} \ge 0.01 \frac{mb}{sr^2\,MeV}$ and this for at least 5~MeV 
along the S-curve ($\sigma$-cut). 
The resulting $\Delta O$'s are shown in the rows labeled $\sigma$-cut.
And finally, in addition to the $\sigma$-cut, we demand that 
the energies of the detected nucleons ($E_1$, $E_2$) must 
be greater than 15~MeV\footnote{3~MeV for $E_{lab}=13$~MeV}. 
The final $\Delta O$'s are displayed in the third rows labeled $\sigma+E$-cut.
Supplementary, 
in the last column of Tab.~\ref{tab:max4scan} labeled $N$ 
we present the  number of configurations fulfilling 
without maximum requirements
the three 
search restrictions:
no restrictions, applying the $\sigma$-cut, and applying both, the 
$\sigma$- and
$E$-cuts. 

The cross section ($\sigma$-cut) 
and energy ($E$-cut) threshold values are based on current 
experimental conditions and limitations. We are aware that each of these 
values is somewhat arbitrary. We do not need to lower them, however, since  
as we show below the 
3NF effects are already clearly visible under these restricted conditions.

As we can see in Tab.~\ref{tab:max4scan} the 3NF effects, in general,   
increase with the energy of the incoming nucleon. 
This tendency also holds  after applying the $\sigma$-cut and the 
$\sigma$-cut together with the $E$-cut.
The only exceptions are $A_x$ (where the maximum value of 
$\Delta A_x$ including the cuts is at $E_{lab}=65$~MeV)  
and $A_y$ and $A_{xx}$ (where the 3NF effects 
are largest at  135~MeV rather than at 200~MeV).
The values of $\Delta O$ after applying both the $\sigma$- and $E$-cuts  
do not change in a considerable way but the number of 
configurations $N$ decreases dramatically.

The polarization observables are particularly 
sensitive to the 3NF. 
Among them the  tensor analyzing powers  
are especially interesting, showing large 3NF effects. 
The nucleon and deuteron vector analyzing powers
($A_{y}$, $A_{y}^{(d)}$) are good tools to test the 3NF as well. 
Predictions for the differential cross section are also very promising,
especially at the energies  135 and 200~MeV.
The 3NF influence is the least evident in  $A_{z}$ and $A_{x}$. 

The next step in our analysis was to divide the phase space 
into regions according to certain values of 
$\delta O=\delta O(\theta_1,\theta_2,\phi_{12},E_1,E_2)$ for our chosen 
observables.  Of course, the regions where $\delta O$ takes its  largest 
values are of the strongest interest. In order to display these 
regions, we project them onto three planes: 
$\theta_1$-$\theta_2$, $\theta_1$-$\phi_{12}$ and $E_1$-$E_2$. 
Fig.~\ref{fig:sigmakolor-przyklad} shows one example for $\Delta\sigma$ 
at 135 MeV. 
The three rows represent the three cut conditions described above and labeled 
as in Tab.~\ref{tab:max4scan}.
 Using the information from the first and  the second column for 
$\Delta\sigma$, varying between 0 and 90\%, one can locate the angular regions 
in phase space where $\Delta\sigma$ values of a certain magnitude can 
be found. The last column provides the additional information in the 
energy distribution. For example our sample plot shows one very promising 
configuration ($(\theta_1\simeq 15^\circ,\theta_2\simeq 15^\circ,
\phi_{12}\simeq 0^\circ)$) with $\Delta\sigma\gtrsim 90\%$. Moreover 
that configuration survives all the applied cuts. 
 The corresponding situation for $A_{yy}$ is shown in 
Fig.~\ref{fig:ayykolor-przyklad}. It is seen that 3NF effects are larger 
in magnitude and they occur in different regions of the phase space compared to the interesting ones for $\Delta \sigma$.

In this way we found at each energy the most promising configurations 
for each studied observable.
 It turned out that the influence of the 3NF is not limited to 
separate points but rather shows up 
in extended regions of phase space. Also the regions differ 
with the observable chosen.

Next we performed more detailed studies. For each observable and each energy 
we have chosen about 20 angular configurations and calculated 
the observable using the five high-precision potentials. In the 
following figures they form the light shaded band. Then 
the TM 3NF was combined with all five NN potentials forming a second band 
(dark shaded band  in the following). 
In addition, we regard the force combinations CD~Bonn with TM' 
and AV18 with Urbana~IX.

As a  result of our study, we have   
a huge amount of information on the 3NF effects 
for each observable at each of the four energies. 
In the following, we present only general 
conclusions about each observable and we only show 
 the most interesting  configurations. 
More results  and examples can be found in ~\cite{phdkuros}.

\subsection{3NF effects at 13~MeV}
\label{sec:bkup13}

\begin{figure}[htbp]
\leftline{\mbox{\epsfysize=20.0cm \epsffile{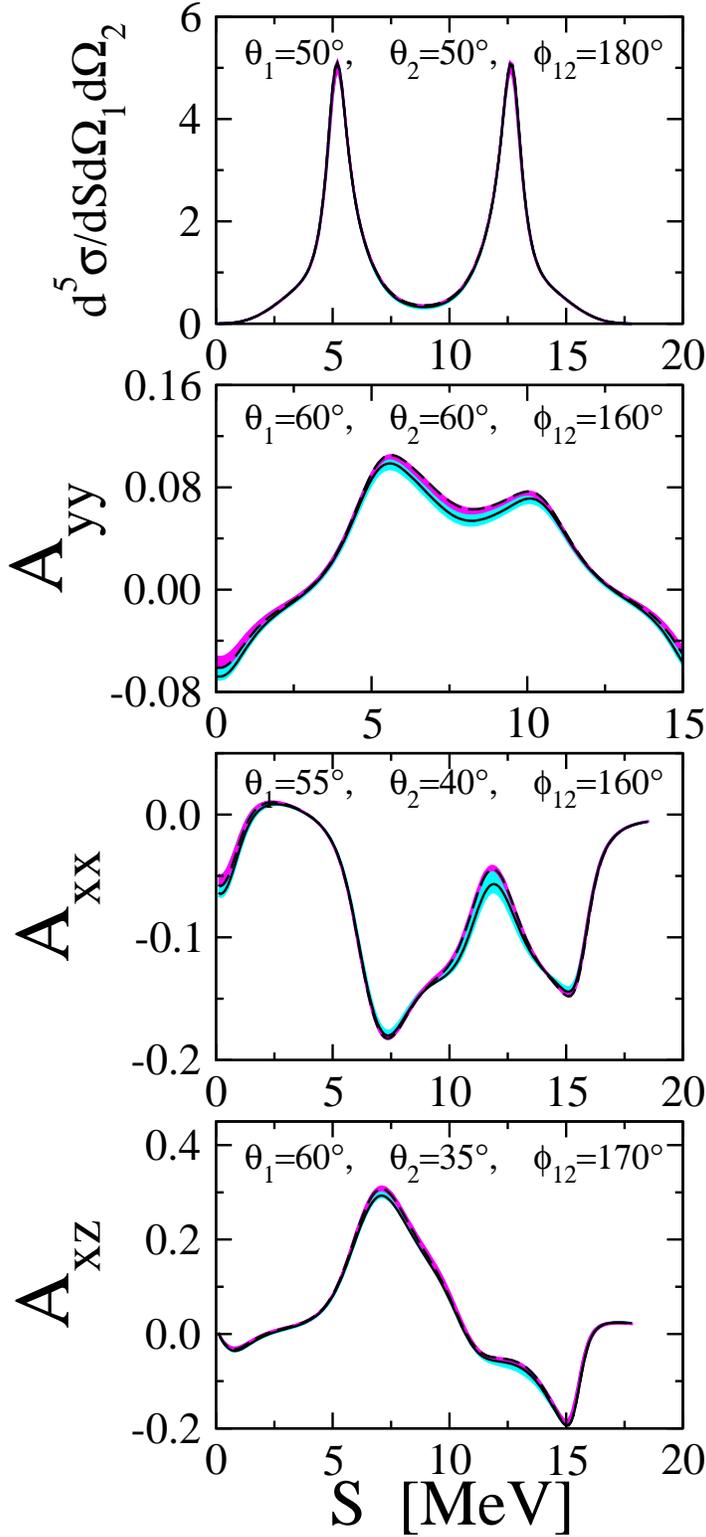}}}
\caption[]{Cross section in [mb MeV$^{-1}$sr$^{-2}$] and analyzing powers for selected breakup 
configurations at 13~MeV. The light shaded band (blue) 
contains  the theoretical predictions based on CD Bonn, AV18, Nijm~I, II 
and Nijm~93. The darker band (red) represents predictions when these 
NN forces are combined with the TM 3NF. The solid line is for AV18+Urbana~IX 
and the dashed line for CD Bonn+TM'. 
}
\label{fig:e13prazemdwj}
\end{figure}

The case of 13~MeV serves as an example for 3NF effects 
in a low energy breakup process.
From Tab.~\ref{tab:max4scan} it follows that for  vector analyzing 
powers  no significant 3NF effects are present. 
The values of $\Delta O$ for tensor analyzing powers and the cross section 
give some hope to find configurations with measurable 3NF effects.
In Fig.~\ref{fig:e13prazemdwj} we show some examples in which we found 
the largest effects 
for the differential cross section, $A_{xx}$, $A_{yy}$ and $A_{xz}$.
Unfortunately, the spread in the  predictions of various 
NN potentials  
can be as large as 10\% and this is   
comparable in magnitude with the  3NF effects  found.
 Therefore the conclusion is that the 3NF effects at 13~MeV  
are too  small to be distinguished from 
the dependences on the NN force model.

\subsection{3NF effects at higher energies} 
\label{sec:details} 

To look  for the magnitude of  3NF effects at higher energies we 
 investigated each of the another eight observables 
listed above at 65, 135 and 200~MeV. 
We  focus only on  configurations 
which survive both the $\sigma$- and  $E$-cuts, 
because our aim is to find regions  of the phase 
space with large and measurable 
  3NF effects. In all cases, however, 
we checked if  the rejected  configurations contain 
 interesting  3NF effects.

\subsubsection{Differential cross section}

At higher energies there  are significant effects of  
3NFs in the differential cross section. 
They are particularly large (more than 90\%) in some configurations 
at 135 and 200~MeV.

\begin{figure}[htbp]
\leftline{\mbox{\epsfxsize=14.0cm \epsffile{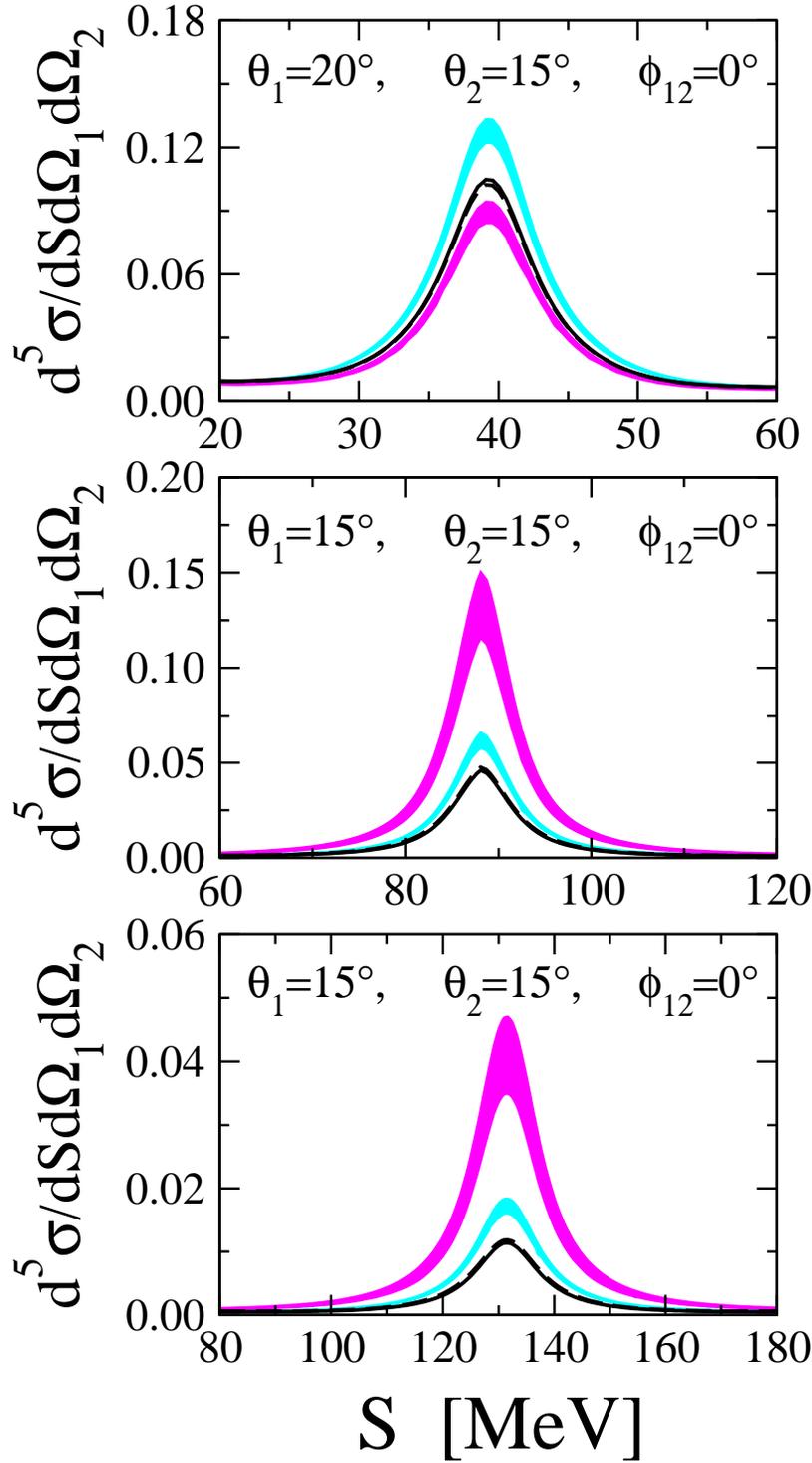}}}
\caption[]{Differential cross section in [mb MeV$^{-1}$sr$^{-2}$] 
in selected breakup configurations 
at 65 (top), 135 (middle)  and 200~MeV (bottom). For the description of bands and lines see 
caption of Fig.~\ref{fig:e13prazemdwj}.}
\label{fig:sigwybrane}
\end{figure}

Generally, in all  phase space points studied 
  adding the TM 3NF increases the value of the cross section at 
135 and 200~MeV, whereas 
at 65~MeV we found  configurations where the TM 3NF decreases $\sigma$. 
We show prominent cases in Fig.~\ref{fig:sigwybrane}.
We see that at 135 and 200 MeV $\delta \sigma \ge 60\%$ 
(in fact this happens not only for $\phi_{12}=0^\circ$ but in a range 
as large as $20^\circ$). In the maxima of the cross sections 
the energies of two outgoing nucleons  ($E_1$ and $E_2$) are comparable 
and together  amount to about 80\% of the total accessible energy. 
The ``2N'' and ``2N+TM'' bands are well separated. 

Astonishingly, 
at 135 and 200 MeV the AV18+Urbana IX and CD~Bonn~TM' predictions, 
while close together, are distinctly different from all 
NN force combinations with TM. 

\subsubsection{Vector analyzing powers $A_y$ and $A_y^{(d)}$}

The 3NF effects for the   nucleon and deuteron 
vector analyzing powers $A_y$ and $A_y^{(d)}$
increase with energy and become quite large 
at higher energies in some specific configurations. 

\begin{figure}[htbp]
\leftline{\mbox{\epsfxsize=14.0cm \epsffile{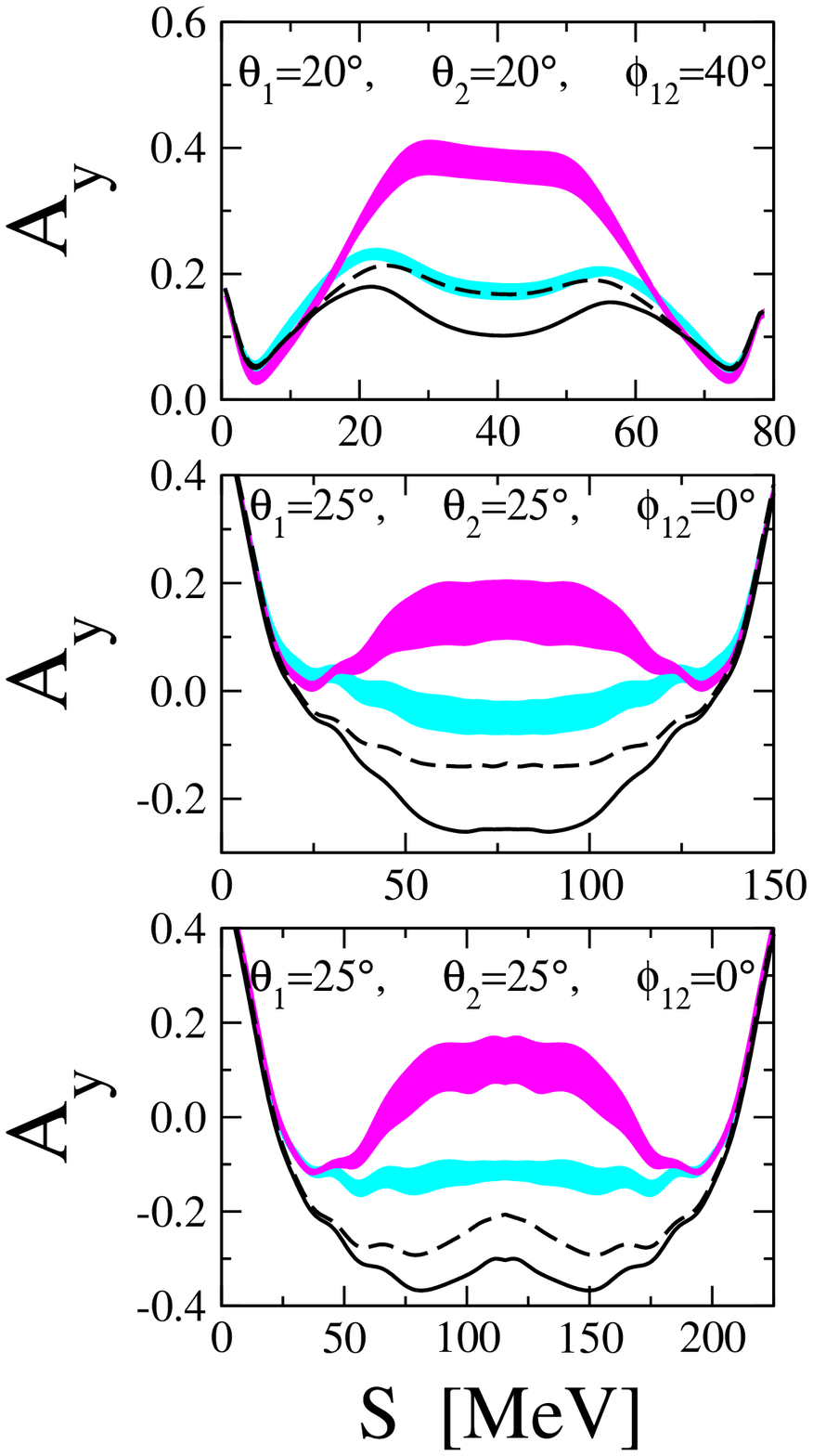}}}
\caption[]{Nucleon  analyzing power  $A_y$ in selected breakup 
configurations at 65 (top), 135 (middle) and 200~MeV (bottom). For the description of 
bands and lines see caption of Fig.~\ref{fig:e13prazemdwj}. }
\label{fig:aynwybrane}
\end{figure}

\begin{figure}[htbp]
\leftline{\mbox{\epsfxsize=14.0cm \epsffile{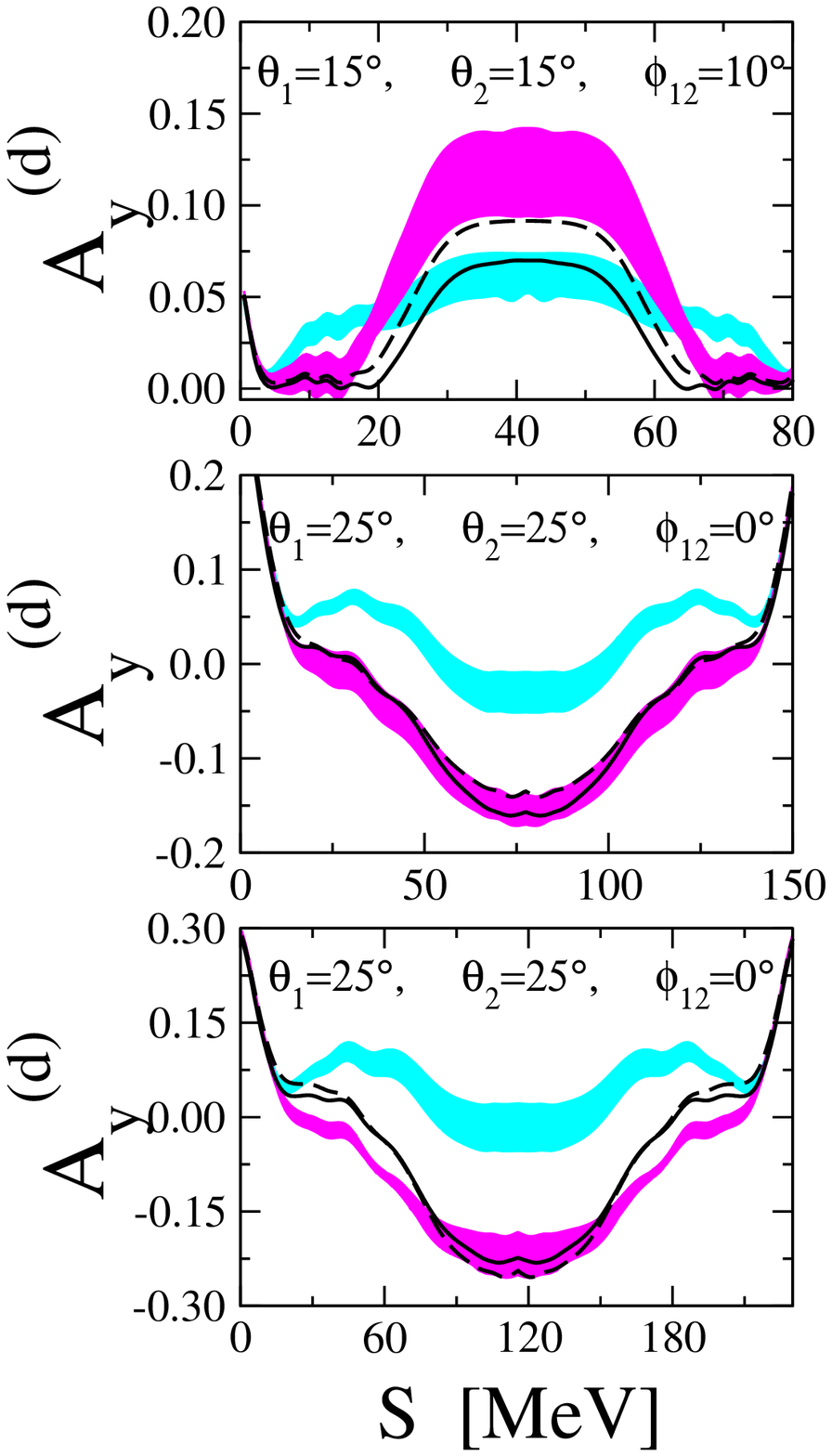}}}
\caption[]{Deuteron vector analyzing power  $A_y^{(d)}$ in selected 
breakup configurations at 65 (top), 135 (middle) and 200~MeV (bottom). For the description of 
bands and lines see caption of Fig.~\ref{fig:e13prazemdwj}. }
\label{fig:aydwybrane}
\end{figure}

Most of these configurations  
are close to  the so-called final state interaction (FSI) geometry,
     where the two outgoing nucleons have equal momenta.
For example, at 200~MeV  
the value of $\delta A_y^{(d)}$ reaches about $350\%$ for the 
($\theta_1,\theta_2,\phi_{12}$)=($25^\circ,25^\circ,0^\circ$) configuration while in others   
$\delta A_y^{(d)} \le 150\%$. 
Similarly, $\delta A_y$ reaches the maximal value of nearly $200\%$, 
but only in few  configurations goes  above 140\% (six for 135~MeV 
and three for 200~MeV).
Figs.~\ref{fig:aynwybrane} and \ref{fig:aydwybrane} show  examples of  
the most interesting configurations at each energy. 
Despite the substantial broadening of both ``2N'' and 
``2N+TM''  bands (especially at 65~MeV), 
 the TM 3NF effects are  clearly visible.
At higher energies the 
predictions with and without  TM 3NF have even different signs, both for 
 $ A_y$ and $ A_y^{(d)}$. We found also some configurations where 
the inclusion of  the TM 3NF  decreases the pure NN value of the 
analyzing powers from the positive 
value 0.2 to  zero (e.g. ($15^\circ,15^\circ,20^\circ$)). 
 Data from such  configurations can be an excellent tool to test  3NF models.

We cannot draw any general conclusions about the 
predictions obtained using Urbana~IX with AV18 and TM' 
with CD~Bonn, since they are quite different in various configurations.
In addition they can be also quite different from the TM predictions.

\subsubsection{Nucleon analyzing powers $A_x$ and $A_z$}

To get nonzero values for the  nucleon  analyzing powers $A_x$ and $A_z$ one
has to go out of the reaction plane~\cite{ref24,Ohlsen2}. 
For both of them the 3NF 
effects are the smallest  among all considered analyzing powers. Moreover,  
the largest  $\delta A_x$  value was found at  65~MeV 
and it decreases with increasing energy (assuming cuts). 
For  $A_z$ 
the maximum values of $\delta A_z$ at 135 and 200~MeV  
are comparable ($\delta A_z \cong 60\% $) 
and they are larger than the corresponding value  
at 65~MeV ($\delta A_z \cong $20\%).
For $A_z$ the assumed threshold 
value of $|A_z|$ ($|A_z|\ge 0.05$) is crucial because    
$|A_z|$ reaches measurable values only at higher energies. 
There this cut is not so decisive anymore because the   
maximum of $|A_z|$ is larger and approaches  0.45. 

\begin{figure}[htbp]
\leftline{\mbox{\epsfxsize=14.0cm \epsffile{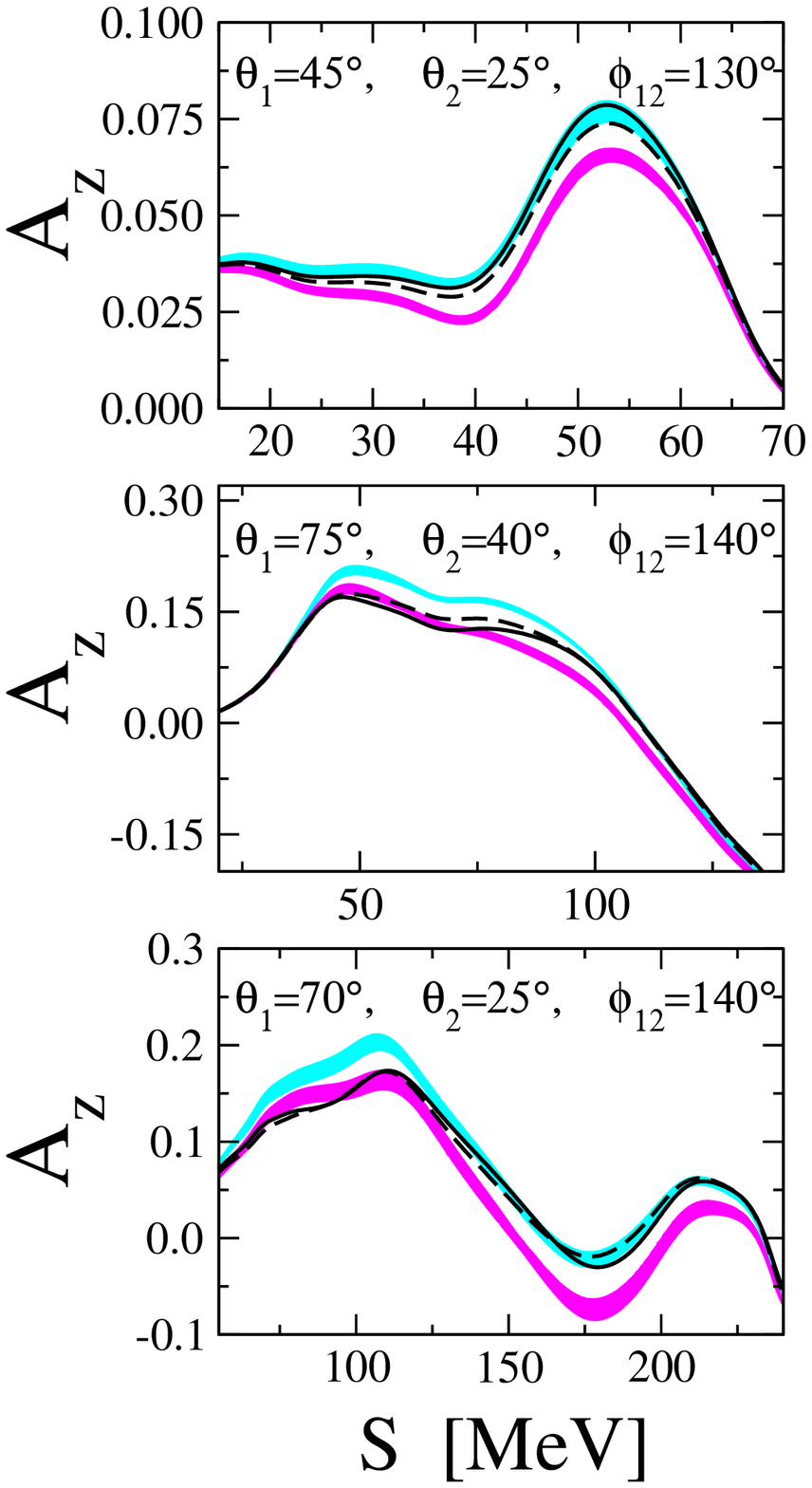}}}
\caption[]{Nucleon  analyzing power  $A_z$ in selected breakup 
configurations at 65 (top), 135 (middle) and 200~MeV (bottom). For the description of bands 
and lines see caption of Fig.~\ref{fig:e13prazemdwj}.}
\label{fig:aznwybrane}
\end{figure}

\begin{figure}[htbp]
\leftline{\mbox{\epsfxsize=14.0cm \epsffile{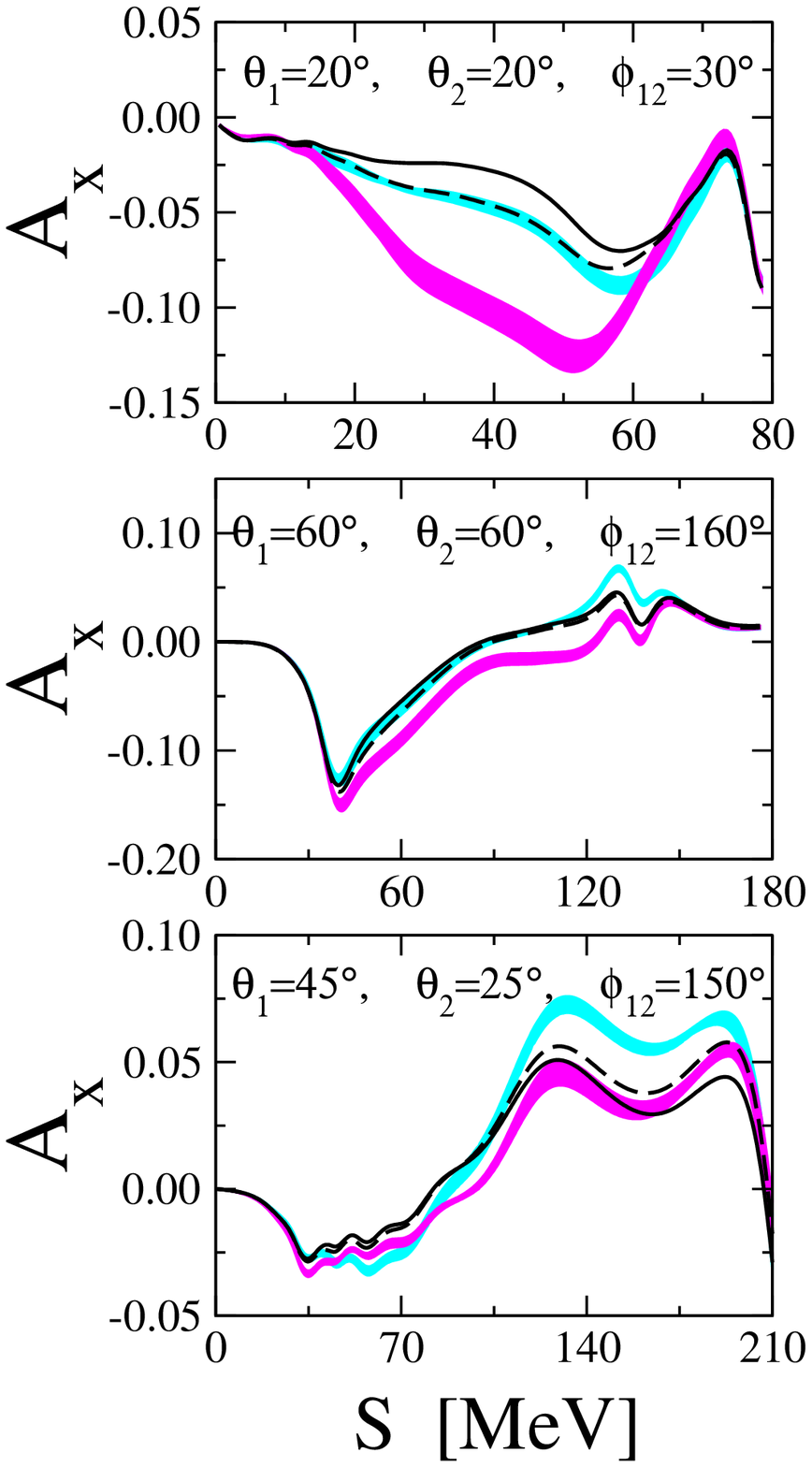}}}
\caption []{Nucleon analyzing power  $A_x$ in selected breakup 
configurations at 65 (top), 135 (middle) and 200~MeV (bottom). For the description of 
bands and lines see caption of Fig.~\ref{fig:e13prazemdwj}. }
\label{fig:axnwybrane}
\end{figure}

Again, we picked up  the most promising configurations and show them in 
Figs.~\ref{fig:aznwybrane} and \ref{fig:axnwybrane} for $A_z$ and $A_x$, 
respectively.
At all energies the  ``2N'' and ``2N+TM'' bands are rather 
narrow and the 3NF effects are 
 clearly visible. 
For $A_x$ and $A_z$ the  Urbana~IX  and  TM' forces  provide  
smaller 3NF effects.  

\subsubsection{Tensor analyzing powers}

\begin{figure}[htbp]
\leftline{\mbox{\epsfxsize=14.0cm \epsffile{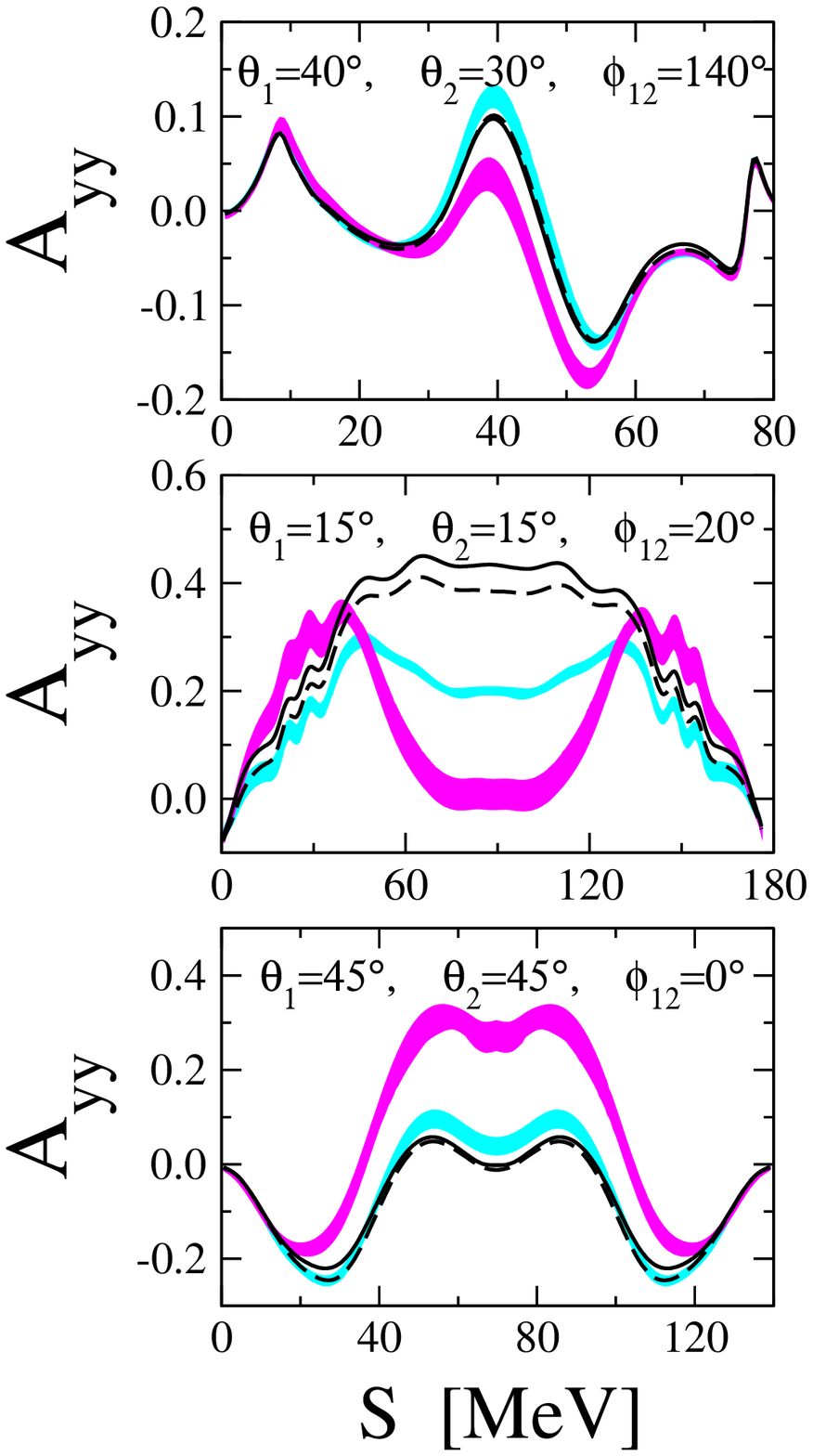}}}
\caption[]{Tensor analyzing power $A_{yy}$ in selected breakup 
configurations at 65 (top), 135 (middle) and 200~MeV (bottom). For the description of 
bands and lines see caption of Fig.~\ref{fig:e13prazemdwj}. }
\label{fig:ayywybrane}
\end{figure}

These spin observables appear to be 
the most sensitive  to the 3NFs we used. 
For each of them the magnitude  of 3NF effects 
increases strongly with the energy.
There are extended regions of the phase space 
with large   3NF effects  at each 
energy, what makes the investigation of  
$A_{xx}$, $A_{yy}$ and $A_{xz}$ extremely promising.
At  the highest energy there are  a few  
configurations with especially large  values of 
$\delta A_i$ ($A_i=A_{yy},~A_{xx},~A_{xz}$). One  distinguished  
example is $A_{yy}$ for which 
$\delta A_{yy}$ reaches the extremely large 
value of about 400\% in the ($45^\circ,45^\circ,0^\circ$) configuration 
(see Fig.~\ref{fig:ayywybrane}). 
For this observable there are   
in addition three configurations with  $\delta A_{yy}\ge 250\% $. 
They  are close to     FSI geometries with comparable 
energies of the two outgoing nucleons. We found a 
similar behavior for $A_{xx}$ and $A_{xz}$.

At the higher energies we even 
found several configurations with  different signs for 
 the ``2N'' and ``2N+TM'' bands  what leads to a very big 
value of $\delta A_i$.
However, they do not  always represent the best region 
  to investigate  3NF effects. 
For example, $\delta A_{xx}$
at 200~MeV reaches the largest value for the 
($145^\circ,10^\circ,180^\circ $) 
configuration  but  the short length  of the S-curve, 
where  3NF effects are large, 
makes this configuration hard to use for an experimental study.
We have also found several configurations with large  3NF effects  but 
the large spreads of ``2N'' and ``2N+TM''  
bands   inhibit the unambiguous conclusion 
about the significance  of 3NF effects, e.g. for $A_{xz}$ at 200~MeV 
in ($45^\circ,45^\circ,0^\circ $) 
or configurations with largest $\delta A_{yy}$. 
On the other hand some of  these configurations present themselves as  
excellent playground  to 
look for  differences between different models of NN interactions. 

\begin{figure}[htbp]
\leftline{\mbox{\epsfxsize=14.0cm \epsffile{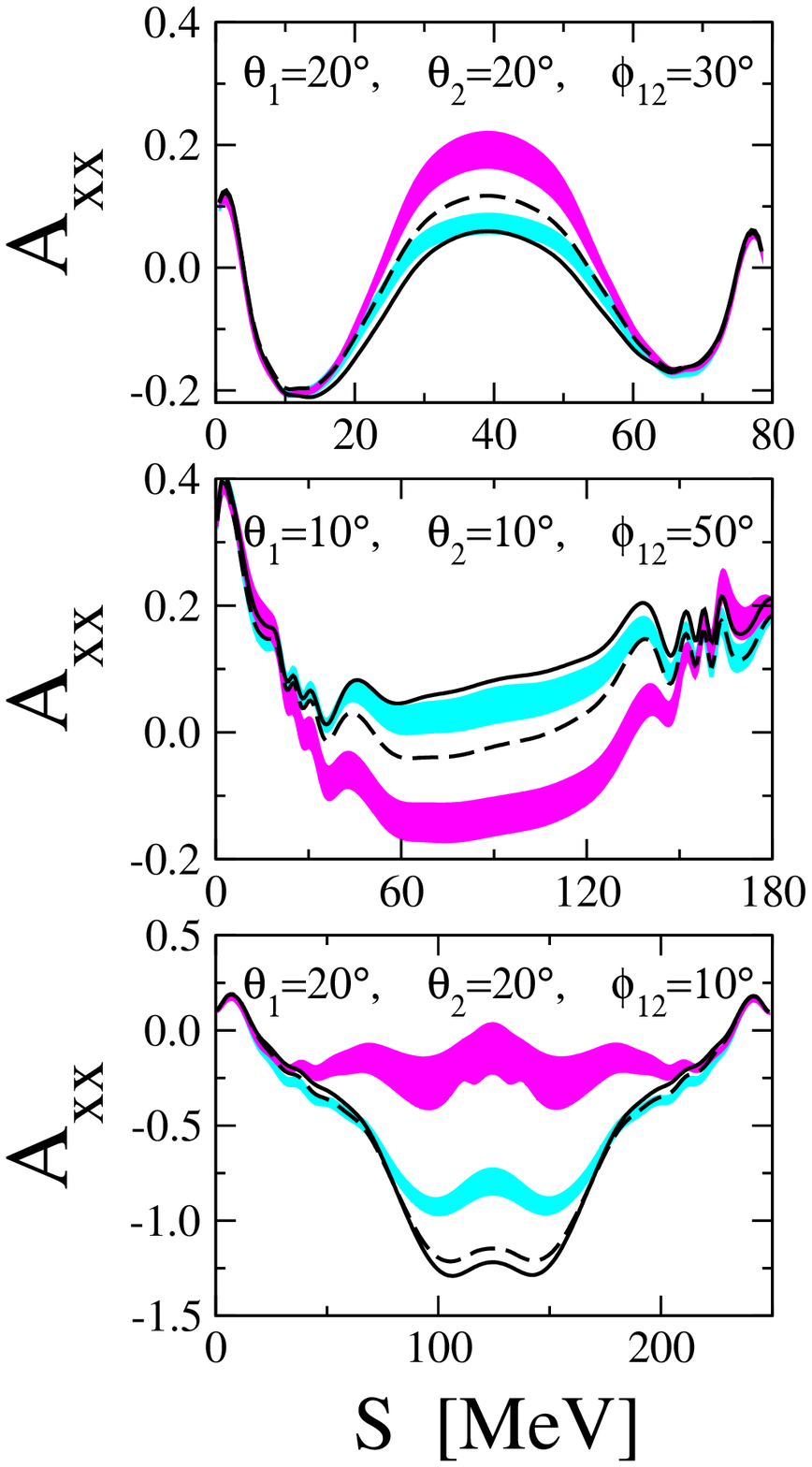}}}
\caption[]{Tensor analyzing power $A_{xx}$ in selected breakup 
configurations at 65 (top), 135 (middle) and 200~MeV (bottom). For the description of 
bands and lines see caption of Fig.~\ref{fig:e13prazemdwj}. }
\label{fig:axxwybrane}
\end{figure}

\begin{figure}[htbp]
\leftline{\mbox{\epsfxsize=14.0cm \epsffile{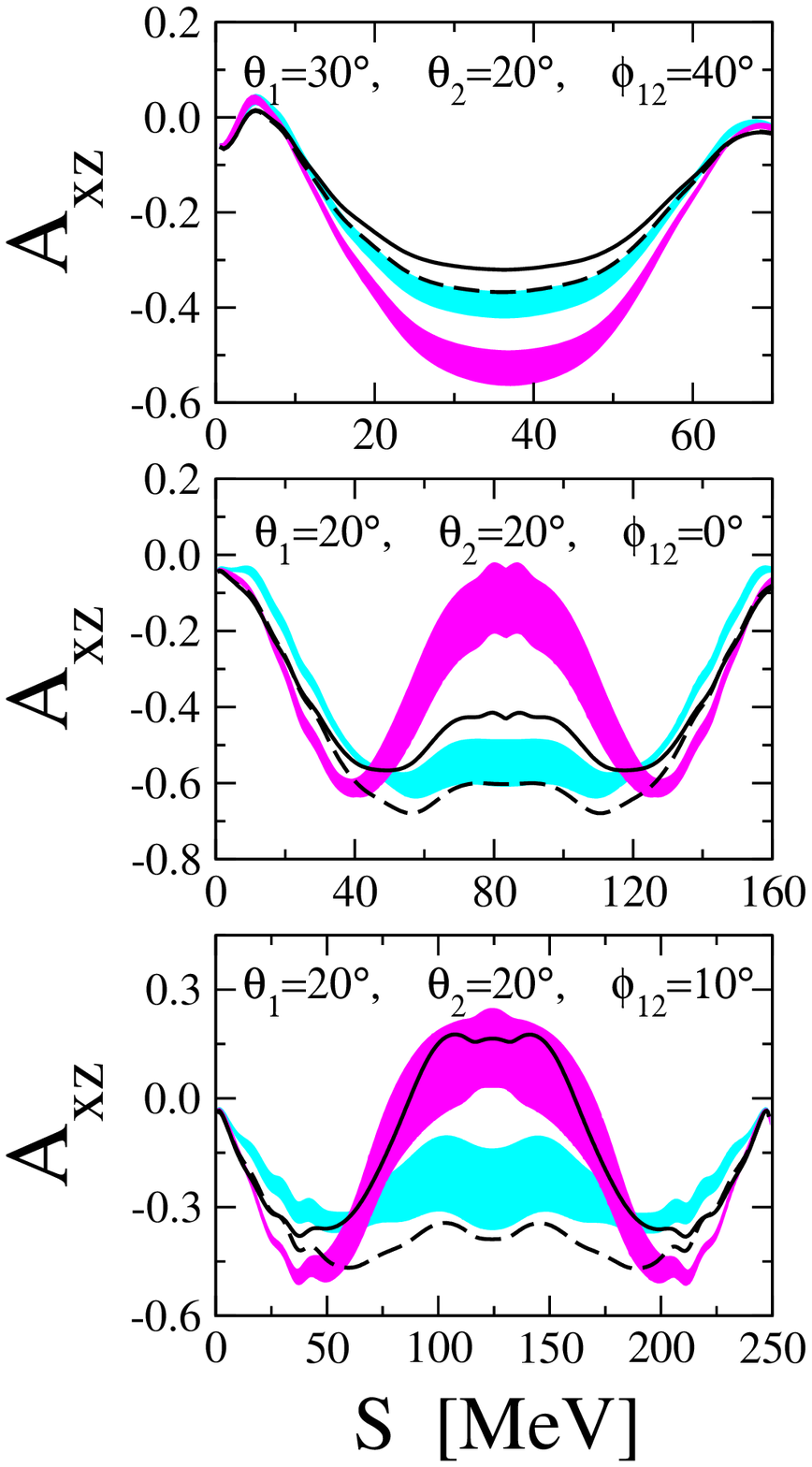}}}
\caption[]{Tensor analyzing power $A_{xz}$ in selected breakup 
configurations at 65 (top), 135 (middle) and 200~MeV (bottom). For the description of 
bands and lines see caption of Fig.~\ref{fig:e13prazemdwj}.}
\label{fig:axzwybrane}
\end{figure}

Figs.~\ref{fig:ayywybrane}, \ref{fig:axxwybrane}  and 
\ref{fig:axzwybrane} show our 
choice for the  most sensitive configurations.
The predictions 
using Urbana~IX 
combined with AV18 and TM' with CD~Bonn are often 
close to each other and  give 
smaller effects   than with the TM 3NF.

\section[]{Search for configurations with 
large 3NF effects in the \bbox{$p(\vec d,pp)n$} breakup at 130~MeV}
\label{secIV}

Awaiting results from a new experiment currently under way at KVI in 
Groningen~\cite{bodek}, we calculated
the differential cross section $\frac{d^5\sigma}{d\Omega_1d\Omega_2dS}$ 
together with  vector and tensor analyzing powers 
$A_{y}^{(d)}$, $A_{xx}$, $A_{yy}$, $A_{xz}$ for the  $p(\vec d,pp)n$ breakup reaction  at 
$E_d=130$~MeV in the laboratory frame. 
All these observables are planned to be measured.
The calculations were performed  for 970 configurations 
covering  the entire experimentally available phase space 
(the  maximal  available angle  of the outgoing protons is 
 $\theta=76.9^\circ$ in the laboratory frame). We used the CD~Bonn potential 
with and without the TM 3NF varying the angles  $\theta_{1}$, $\theta_{2}$  
and $\phi_{12}$ in steps of $5^\circ$ and the position 
on the kinematical locus $S(E_1,E_2)$ in steps of 1~MeV. 
This exploratory study with the only force combination CD~Bonn and TM
was performed  in the same way as in the previous 
Section~\ref{secIII} and
the  results  are presented in Tab.~\ref{tab:e65d-max}.
Again the tensor analyzing powers are the most sensitive observables 
to 3NFs. Then we performed in addition a more detailed study
using all force combinations as in the previous Section.

\subsection{Differential cross section}

In the case of the differential cross section the experimentally required condition  
$E_i \ge 15$~MeV  rejects 
a lot of very promising configurations with $\phi_{12} \le 70^\circ$. 
 Despite that, we found many configurations with 3NF effects  
in the region of the maximal values (see Tab.~\ref{tab:e65d-max}). 
 Very interesting configurations are 
$(\theta_1\simeq 20^\circ,\theta_2\simeq 20^\circ,\phi_{12} \ge 130^\circ)$
which correspond to the situation where  one of the outgoing 
nucleons takes all accessible energy. 

\begin{figure}[htbp]
\leftline{\mbox{\epsfxsize=14.0cm \epsffile{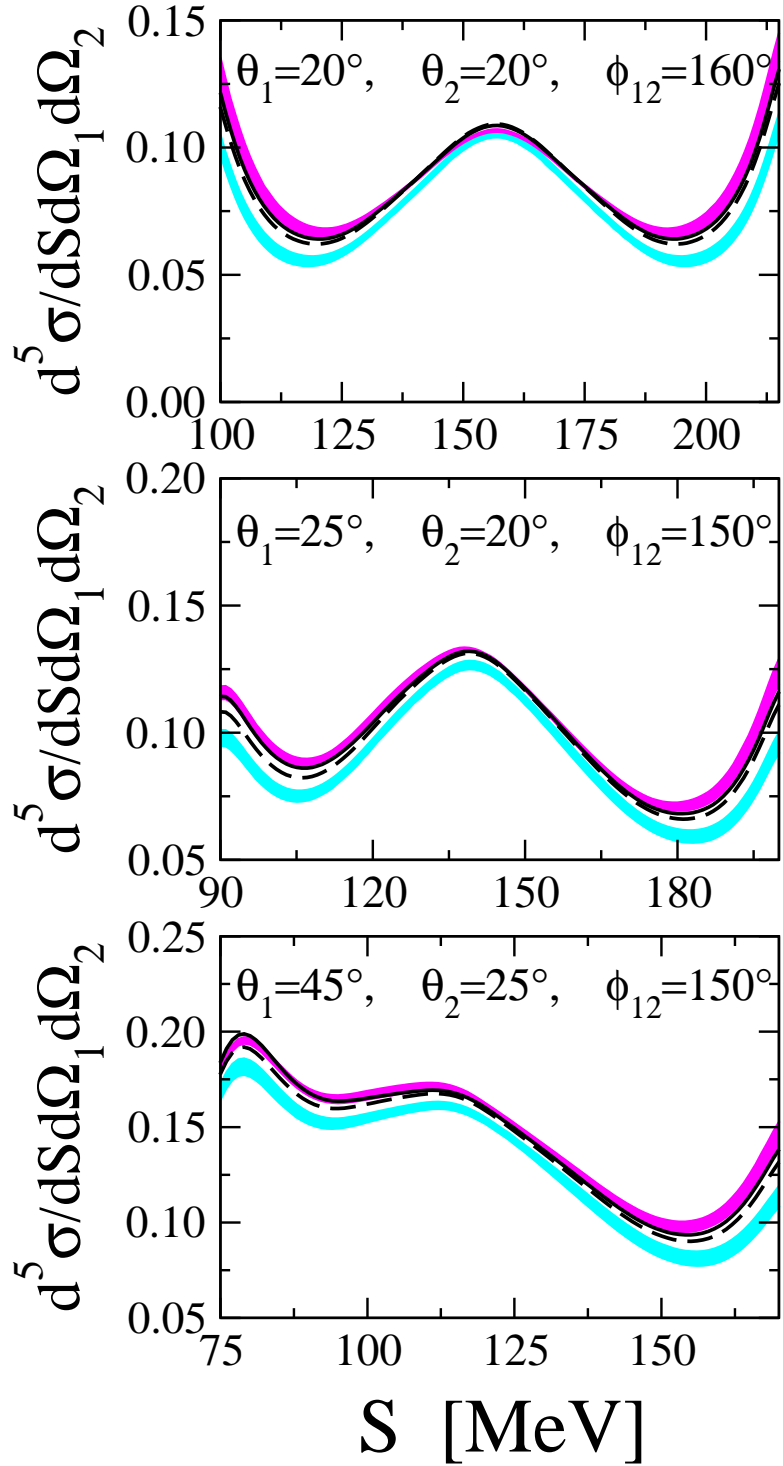}}}
\caption[]{Theoretical predictions for the cross section in [mb MeV$^{-1}$sr$^{-2}$] of 
the  dN breakup  at 130~MeV for selected configurations.
For the description of bands and lines see caption 
of Fig.~\ref{fig:e13prazemdwj}.
The points along the  $S$-curve displayed in the figures correspond to the 
ones accessible in the Groningen experiment.}
\label{fig:e65dsigwybrane}
\end{figure}

In Fig.~\ref{fig:e65dsigwybrane} we show  cross sections for three
 configurations revealing one of the largest 3NF effects.
Only part of the allowed S-curve is presented, for which 
the energies of  the protons  fulfill the conditions $E_{1,2} \ge 15$~MeV.  
It is very encouraging that 
all of these configurations are accessible in the Groningen experiment. 
The  NN force predictions are rather close together. 
Including the TM 3NF generally increases  the cross section in these  configurations. 
Predictions based on the TM' and Urbana~IX forces are of similar magnitude to the TM  ones. 

\subsection{Deuteron vector analyzing power}
\begin{figure}[htbp]
\leftline{\mbox{\epsfxsize=14.0cm \epsffile{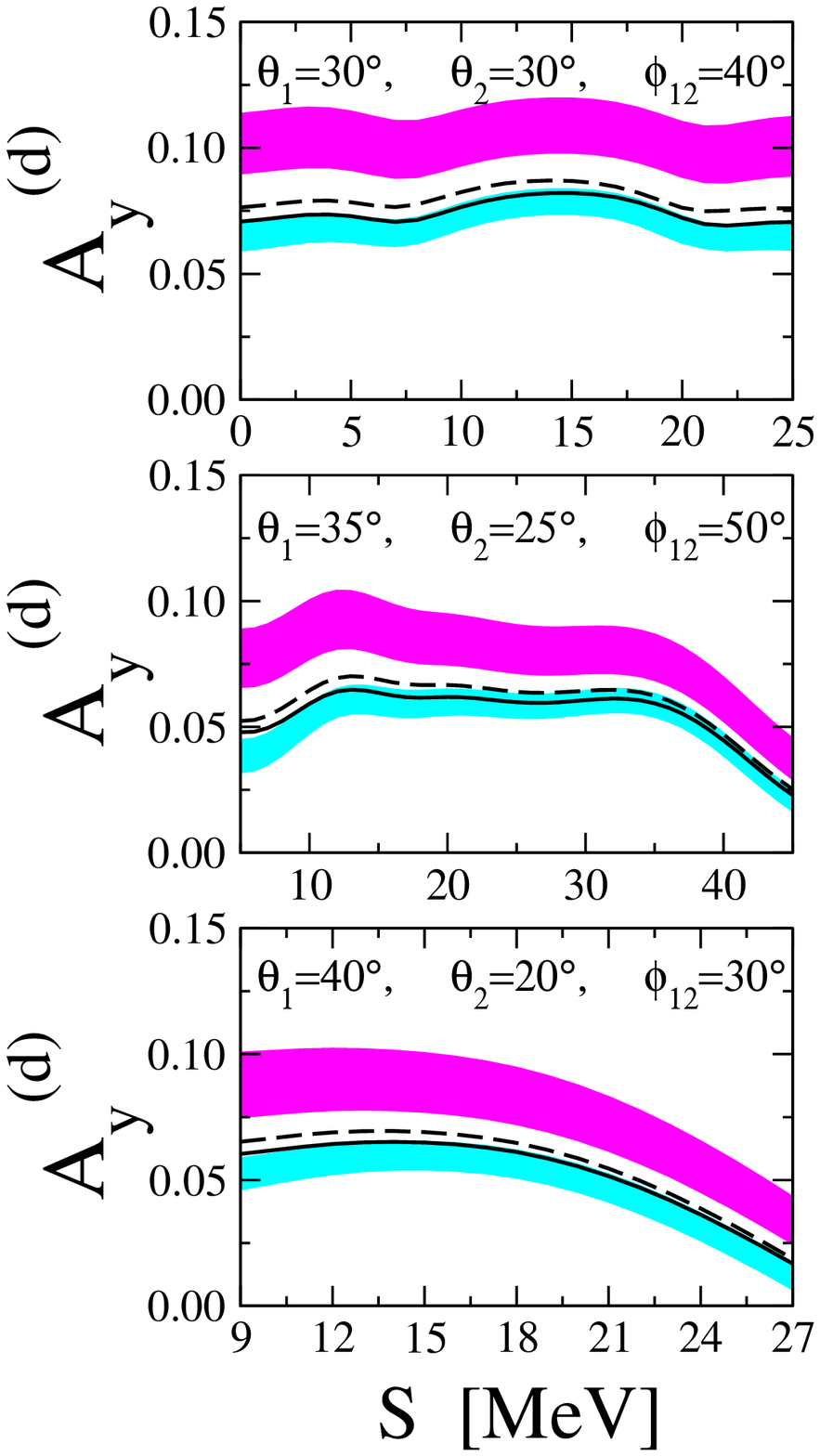}}}
\caption[]{Theoretical predictions for the deuteron vector analyzing 
power $A_y^{(d)}$ of the dN breakup  at 130~MeV for selected 
configurations. For the description of bands and lines see 
caption of Fig.~\ref{fig:e65dsigwybrane}. }
    \label{fig:e65daydwybrane}
\end{figure}

The $A_{y}^{(d)}$ seems to be a  good tool to investigate 3NF effects. 
The most sensitive  configurations occur where the 
energies of two outgoing protons are small ($< 35$~MeV). Unfortunately due to 
the experimental arrangement, which  allows to measure only protons with 
$E_i \ge 15$~MeV many interesting configurations will not be 
studied in the experiment. 
 Nevertheless we found  extended regions of  the phase space 
with  $\theta_{1,2} \ge 20^\circ$ and  
$\phi_{12} \le 50^\circ$  where the value of $\delta A_{y}^{(d)}$ reaches 
more than 60\%. 
From these regions we picked a few configurations and show them 
in Fig.~\ref{fig:e65daydwybrane}.  
The ``2N'' and ``2N+TM''  bands 
are well separated  showing clear and large (up to 70\%) 
3NF effects.  
The NN force band is rather narrow, whereas the ``2N+TM'' 
band is distinctly broader.
For the chosen configurations the CD~Bonn+TM'  and AV18+Urbana~IX  
are close to each other and predict very small 
3NF effects, different from that of the TM 3NF.

\subsection{Tensor analyzing powers}

\begin{figure}[htbp]
\leftline{\mbox{\epsfxsize=14.0cm \epsffile{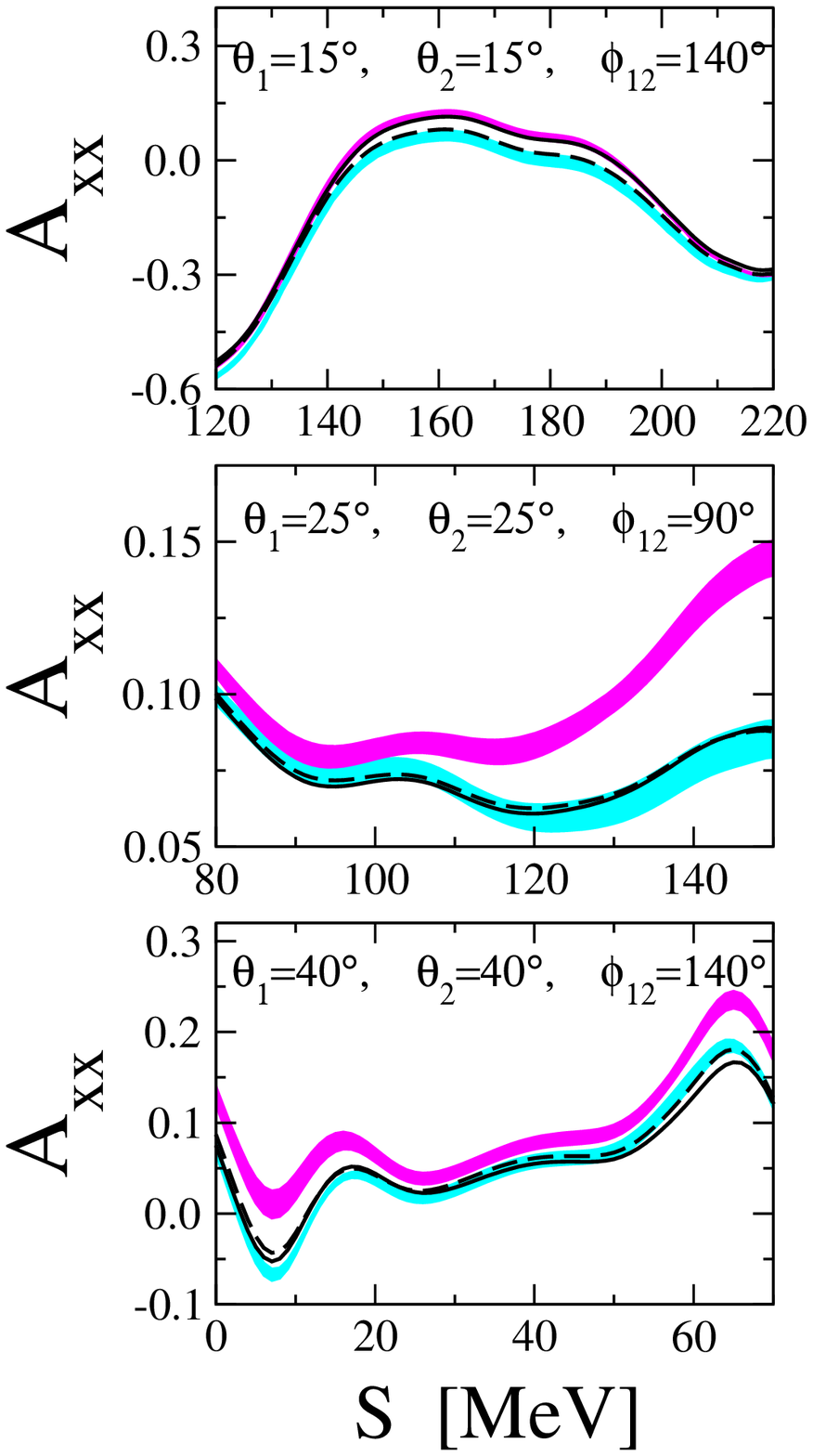}}}
\caption[]{Theoretical predictions for the tensor analyzing power 
$A_{xx}$ of the  dN breakup  at 130~MeV for selected configurations. 
For the description of bands and lines see caption of 
Fig.~\ref{fig:e65dsigwybrane}.}
\label{fig:e65daxxwybrane}
\end{figure}

Finally, we consider the three tensor analyzing powers 
$A_{xx}$, $A_{yy}$ and $A_{xz}$. 
The effects in $A_{xx}$  strongly  depend on the energy 
and, unfortunately, the most interesting part  
of the phase space with largest effects  is outside  
the range of the Groningen experiment 
($E_{1,2} \le 15$~MeV and $ \theta_{1,2} \ge 40^\circ$). 
However, we found  some regions where smaller  but 
clear  3NF effects are  visible 
(see Fig.~\ref{fig:e65daxxwybrane}).
The predictions obtained using Urbana~IX and TM' forces are close together 
and often they do not  differ from the pure NN force predictions.

\begin{figure}[htbp]
\leftline{\mbox{\epsfxsize=14.0cm \epsffile{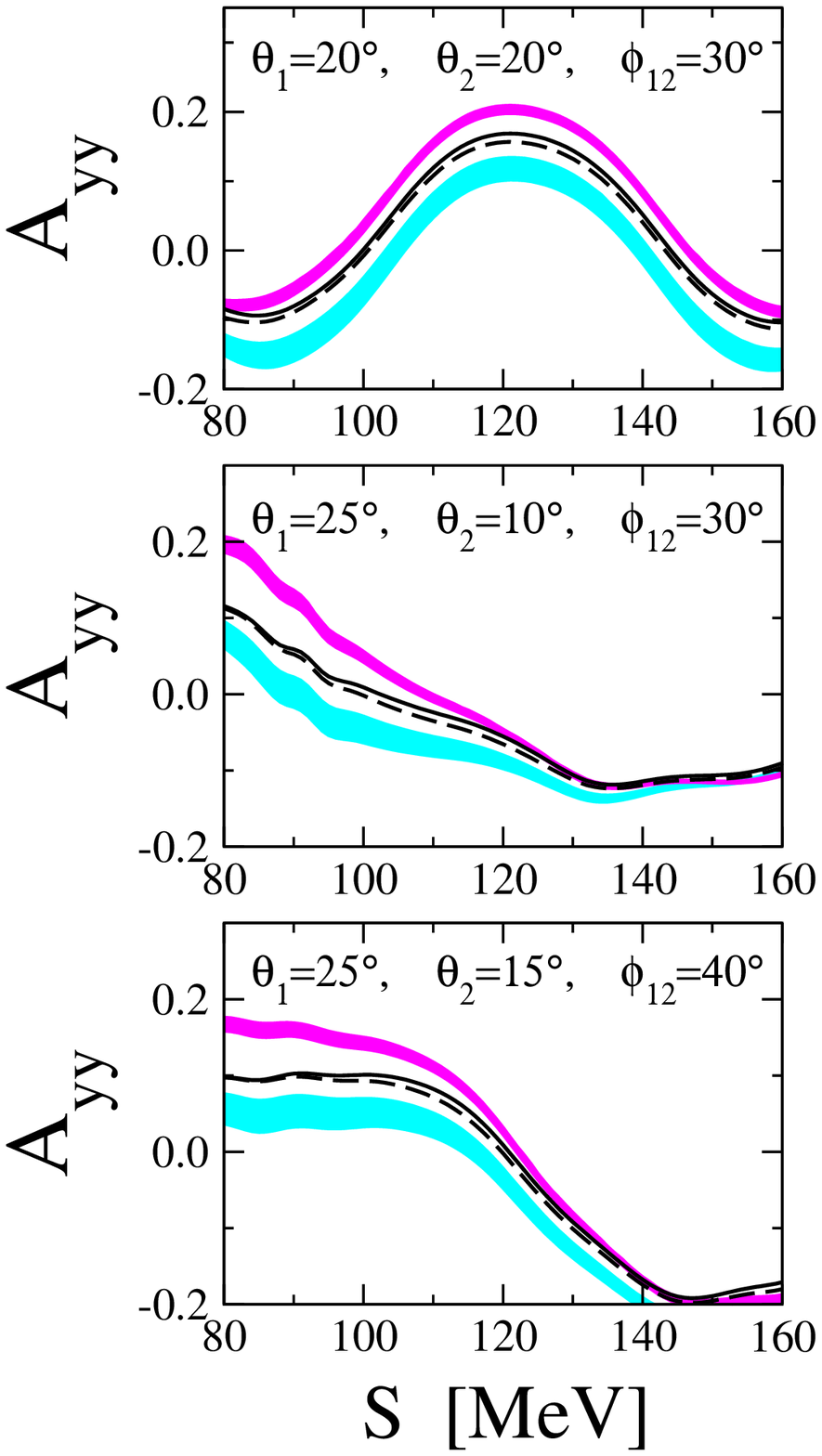}}}
\caption[]{Theoretical predictions for the tensor analyzing power 
$A_{yy}$ of the dN  breakup  at 130~MeV for selected configurations. 
For the description of bands and lines see caption of 
Fig.~\ref{fig:e65dsigwybrane}. }
\label{fig:e65dayywybrane}
\end{figure}

For  $A_{yy}$ 
we found that the most sensitive parts of the 
accessible  phase space are for 
$20^\circ \le \theta_1 \le 25^\circ$ and 
$\theta_2 \le 25^\circ$ with $\phi_{12} \le 60^\circ$.
In Fig.~\ref{fig:e65dayywybrane} we show the three most interesting 
configurations. 
The ``2N'' and ``2N+TM'' bands are rather narrow and 
large  3NF effects rising above  100\% 
are  clearly visible. 
The Urbana~IX and TM' predictions  are rather close together 
and give   smaller 
effects than the TM 3NF.

\begin{figure}[htbp]
\leftline{\mbox{\epsfxsize=14.0cm \epsffile{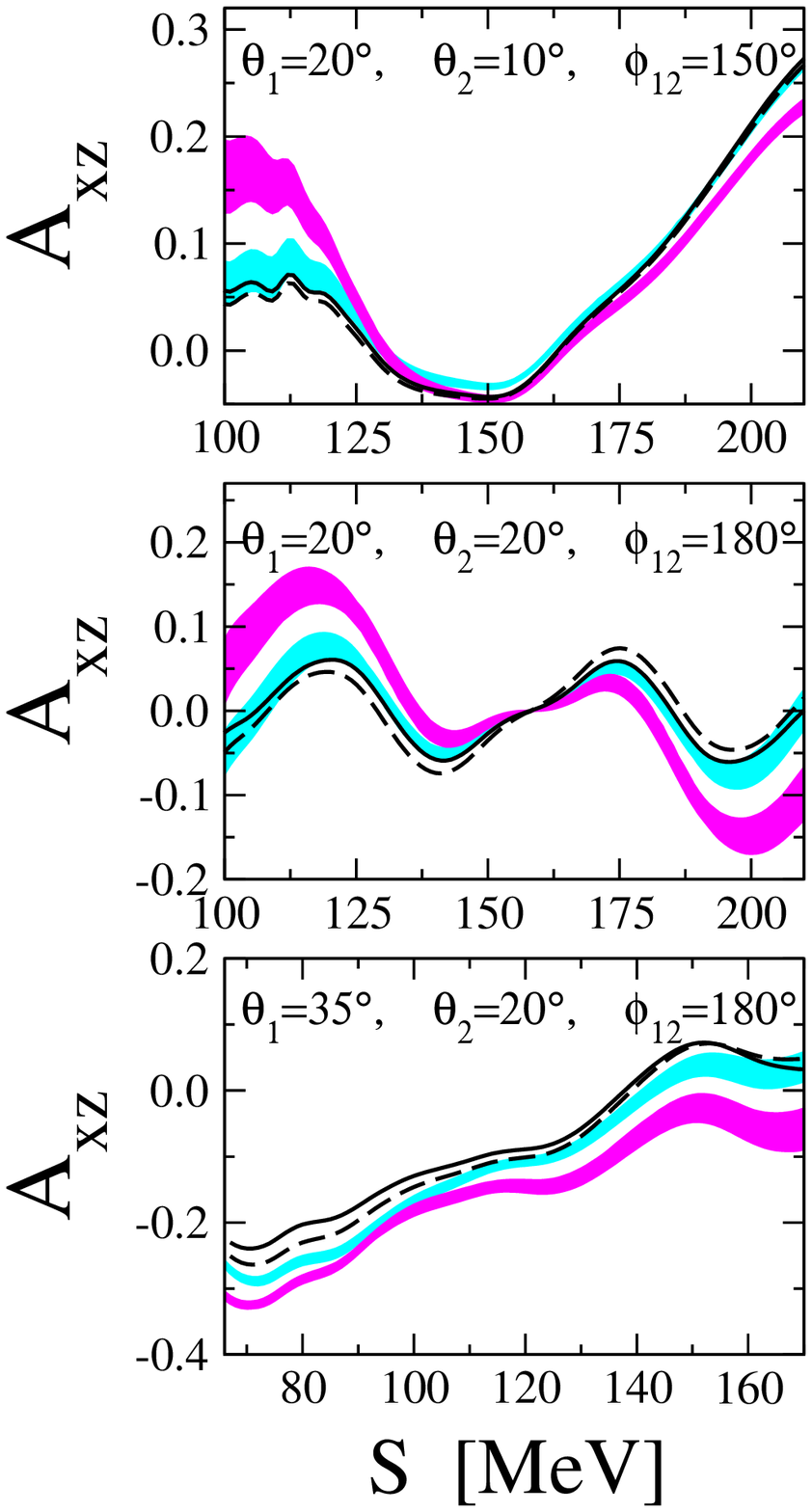}}}
\caption[]{Theoretical predictions for the tensor analyzing power 
$A_{xz}$ of the dN breakup  at 130~MeV for selected configurations. 
For the description of bands and lines see caption 
of Fig.~\ref{fig:e65dsigwybrane}. }
\label{fig:e65daxzwybrane}
\end{figure}

For $A_{xz}$ the configurations with  largest effects 
are for $\theta_{1,2} \cong 20^\circ$ and $\phi_{12} \ge 130^\circ$. 
These configurations correspond to the situation where the  
energy of one of the outgoing 
nucleons is  much larger than the energy of the other nucleon.
In Fig.~\ref{fig:e65daxzwybrane} we show three examples where the 3NF effects 
are larger then 100\%.
Again the effects  of Urbana~IX and  TM' are 
significantly smaller than the effects of TM 3NF.

In conclusion, in the  $\vec d$N breakup reaction 
with an incoming deuteron energy of 130~MeV 
the 3NF effects are clearly visible for cross section,  vector and 
tensor analyzing powers in numerous configurations. 
The most promising observables are the tensor analyzing powers
where the 3NF effects are as large as 100-180\% .
For these observables the magnitudes of the effects depend 
on the 3NF model used.

\section{Summary and conclusions}
\label{secV}
  We started our investigation with the $nd$ total cross section. As could be
foreseen we found that the magnitude of 3NF effects increases with energy.
Because at higher energies the breakup contribution to the total $nd$ cross
section becomes dominant this implies that  the breakup process should
be especially gratifying to see 3NF effects. In such a process, where the
 three final momenta and, in addition, also some spin orientations are fixed,  
one has
the most sensitive insight into the 3N Hamiltonian. Therefore we performed a
detailed 
study
of the $nd$ breakup reaction focusing our investigations on
the effects of 3NFs on the cross section
and quite a few polarization observables.
We  compared  predictions based  on  realistic NN forces
(CD~Bonn, AV18, Nijm~I, II and 93)
to those obtained  when, in addition,
three different  3NF models (TM, TM' and Urbana IX) were included in
the 3N  Hamiltonian.
The NN
force predictions only have a certain spread. Also adding 3NFs to them
yields again a spread. We talk of 3NF effects if the two spreads are clearly
distinct. 
We have found in cross sections and  analyzing powers
large  3NF effects which clearly should be discernible by experiments.

In our study we covered the entire phase space at several energies:
13, 65, 135 and 200 MeV. We
located for each observable the regions in phase space where the largest 3NF
effects occur. These regions are in general different for the various
observables. The 3NF effects increase in general with energy. Especially
promising will be the tensor analyzing powers
$A_{xx}$, $A_{yy}$ and $A_{xz}$.
Also the five-fold
differential cross section  turned out to be a promising observable  to clearly see 3NF effects. 
The smallest effects have been found for
the analyzing powers $A_x$ and $A_z$.
We did not study all possible force combinations between the five high precision
NN forces and the three 3NF models,  which are currently most
often used. Our study aimed only at a first orientation. But the results
show that the outcome for the 3NF effects can depend very much on the force
combination. This tells that the 3NF effects predicted should be considered
with some caution. All the forces are still widely based on phenomenology.
Nevertheless the results indicate that  very likely the  effects will not be
distributed rather smoothly over the whole phase space but  the large
effects will be
concentrated in special regions. Our study might therefore help to
guide future experimental searches.

As an example of such guidance we performed calculations for the
$p(\vec d,pp)n$
experiment currently under way at KVI Groningen. This experiment
covers a good fraction of the full phase space. It will be very interesting
to see how the upcoming data for cross section, vector and tensor analyzing
powers will be related to our predictions.

We have just begun to understand the role which  3NFs play in a nuclear 
Hamiltonian. 
Our present study, though  not complete, clearly shows  the significance and 
importance of  the 3NF effects in the breakup process, 
especially at higher energies. 
A natural extension of this study is the four-nucleon scattering that 
might reveal even more sensitivity to the underlying forces. The first steps 
toward that goal have 
already been made~\cite{viv98,fon99,cies99}.

In order to fully understand the Nd scattering dynamics at 
higher energies 
it is necessary,  in addition to the inclusion of  3NFs, to   
formulate the equations in a 
relativistically covariant  way. 
Also  exact inclusion of the $pp$ Coulomb force in $pd$ scattering  
 at energies above the deuteron breakup threshold is still unsolved and 
leaves a very uncomfortable theoretical uncertainty  
in the analysis of $pd$ data. 
The solution of these two  problems: 
the relativistic formulation of 3N scattering equations  and 
the exact inclusion of the Coulomb 
interactions above the deuteron breakup threshold 
will remove the remaining uncertainties 
and will  make Nd scattering an extremely precise
tool for testing nuclear interactions. It will also  allow to compute exact 3N 
wave functions that can be used to study 
electromagnetic processes in the 3N system
for  a wide range of energies.  

\acknowledgements

This work was supported by
the Polish Committee for Scientific Research (Grant No. 2P03B02818 and 5P03B12320),
 the Deutsche Forschungsgemeinschaft (J.G.), and the NSF (Grant No. PHY0070858). 
One of us (W.G.) would like to thank the Foundation for Polish Science
for the financial support during his stay in Cracow.  
R.S. is a holder of scholarship from the Foundation for Polish Science
and acknowledges financial support.
The numerical calculations have been performed on the Cray T90 and T3E of the
NIC in J\"ulich, Germany.


\begin{table}
\leavevmode
\caption[Comparison of 
$\sigma_{tot}^{br}$ with 
$\sigma_{tot}^{scan}$.]{Comparison of the 
$nd$ total breakup cross section  
$\sigma_{tot}^{br}$ with the value 
$\sigma_{tot}^{scan}$ evaluated according to 
Eq.(\ref{eq:scan}) at four energies of the incoming nucleon.}
\begin{center}
\begin{tabular}{|c|r|r|}
\hline
$E_{lab}$  [MeV]&$\sigma_{tot}^{br}$  [mb]&$\sigma_{tot}^{scan}$  [mb]\\
 \hline
13&167.5&159.2\\
65&91.7&87.3\\
135&56.0&53.5\\
200&48.2&45.9\\
\hline
\end{tabular}
\end{center}
\label{tab:br-scan}
\end{table}

\begin{table}[htpb]
\leavevmode
\caption[$\Delta O $ for eight studied observables and four energies (with 
and without restrictions).]{The maximal changes  
$\Delta O = \max [\delta O(E_{1}, E_{2},\theta _{1}, \theta _{2},\phi_{12})]$ 
due to 3NF effects
for the eight studied observables. For each energy the first row, 
labeled all, gives  $\Delta O$ without constraints,
the second row, labeled $\sigma$-cut, gives  $\Delta O$ under the condition
$\sigma^{(2N)} \ge 0.01 \frac{mb}{sr^2\,MeV}$,
the third row, labelled  $\sigma + E$-cut, gives  $\Delta O$ 
under the additional restriction that 
 $E_{1,2}\ge 15 (3) $~MeV for $E_{lab}$= 65, 135, 200 (13) MeV.
The last column gives the total number of studied configurations (covering 
all values of  $\Delta O$).} 
\begin{center}
\begin{tabular}{|c|r|c|c|c|c|c|c|c|c||c|}
\hline
$E_{lab}$&cuts&$\Delta \sigma$&$\Delta A_{x}$&$\Delta A_{z}$&$\Delta A_{y}$&$\Delta A_{y}^{(d)}$&$\Delta A_{xx}$&$\Delta A_{yy}$&$\Delta A_{xz}$&$N$\\ 
$[$MeV$]$  &    &   [\%]        &  [\%]        & [\%]         &[\%]           & [\%]           & [\%]         & [\%]          &[\%]           &\\
 \hline
$13$  & all           &10 &   0 &   0 &   7 &  10 &  22 &  32 &  19 &6147\\
      &$\sigma$-cut   &10 &   0 &   0 &   7 &  10 &  22 &  32 &  19 &6008\\
      &$\sigma+E$-cut        &10 &   0 &   0 &   0 &   0 &  18 &  17 &  15 &1532\\
 \hline
$65$  & all           &23& 110 & 33 & 182 &  88 & 188 & 194 & 211 &7893\\
      &$\sigma$-cut   &23&  97 & 33 & 156 &  88 & 188 & 194 & 205 &7323\\
      &$\sigma+E$-cut        &23&  97 & 22 & 156 &  88 & 188 & 126 & 130 &2002\\
 \hline
$135$ & all           &98& 202 & 99  & 327 & 289 & 357 & 281 & 577 &8253\\
      &$\sigma$-cut   &98&  93 & 58  & 269 & 233 & 351 & 166 & 195 &7370\\
      &$\sigma+E$-cut        &98&  76 & 58  & 269 & 233 & 351 & 154 & 195 &1028\\
 \hline
$200$ & all           &133& 251 & 205 & 418 & 458 & 580 & 702 & 677 &8597\\
      &$\sigma$-cut   &115&  81 &  60 & 238 & 368 & 275 & 443 & 230 &6638\\
      &$\sigma+E$-cut        &115&  59 &  60 & 238 & 368 & 275 & 443 & 230 & 689\\
\hline
\end{tabular} 
\end{center}
\label{tab:max4scan}
\end{table}

\begin{table}[htbp]
\caption[$\Delta O $ for five studied observables in $p(\vec d,pp)n$ breakup at 130~MeV (with and without experimental limitations).]{The same as in Table~II for five observables in the $p(\vec d,pp)n$ breakup reaction at 130~MeV.}
\leavevmode
\begin{center}
\begin{tabular}{|c|c|c|c|c|c||c|}
\hline
 cuts&$\Delta \sigma$&$\Delta A_{y}^{(d)}$&$\Delta A_{xx}$&$\Delta A_{yy}$&$\Delta A_{xz}$&N\\
 \hline
 all          & 23 & 85 & 197 & 181 & 172 & 970 \\
 \hline
$\sigma$-cut& 23 & 85 & 197 & 181 & 172 & 970 \\
 \hline
$\sigma+E$-cut     & 21 & 80 & 100 & 181 & 172 & 955 \\
\hline
\end{tabular} 
\end{center}
\label{tab:e65d-max}
\end{table}


\thispagestyle{empty}

\newpage

\newpage
\newpage
\newpage

\newpage

\newpage

\begin{figure}[hp]
\caption[]{Two-dimensional projections of phase space regions 
for $\Delta\sigma$ in  the Nd breakup at 
135~MeV. In the first row (all)  phase space points are plotted without  any 
restrictions, in the second row ($\sigma$-cut), and the 
third row ($\sigma + E$-cut) the remaining points are plotted after 
the corresponding cuts 
have been performed. The three columns show  
the $\theta_{1}-\theta_{2}$,  
$\theta_{1}-\phi_{12}$ and $E_{1}-E_{2}$ planes, respectively. 
The different colours express the $\Delta \sigma$ values in percentage.}
\label{fig:sigmakolor-przyklad}
\end{figure} 

\newpage

\begin{figure}[hp]
\caption[]{Two-dimensional projections of phase space regions 
for $\Delta A_{yy}$ in  the Nd breakup at 
135~MeV. For the description  see caption of Fig.~\ref{fig:sigmakolor-przyklad}. }
\label{fig:ayykolor-przyklad}
\end{figure}

\newpage

\thispagestyle{empty}

\end{document}